\pdfoutput=1

\documentclass[%
peerreview%
]{IEEEtran}

\usepackage[square,numbers,sort&compress]{natbib}
\usepackage{hyperref} %
\usepackage%
{hyperref} %
\hypersetup{
    colorlinks,
    linkcolor={blue!80!black},%
    citecolor={green!50!black},
    urlcolor={blue!80!black}
}

\usepackage{amsthm}
\usepackage{amsmath}
\usepackage{mathrsfs}
\usepackage{amssymb} 
\usepackage{stackrel}
\usepackage{mathtools}

\usepackage{physics}

\usepackage{verbatim}
\usepackage{enumerate}

\usepackage{slashbox} 

\usepackage[T1]{fontenc}
\usepackage{bbm} 
\usepackage{bbding}
\usepackage{keystroke}
\usepackage{pifont}
\usepackage{relsize} %

\usepackage{graphicx}
\usepackage{xcolor}
\usepackage{tikz}
\usetikzlibrary{calc}
\usepackage{tabularx}
\usetikzlibrary{arrows,shapes}

\renewcommand{\trace}{\mathrm{Tr}}

\newcommand{\identity}{\mathbbm{1}}

\newcommand{\channel}{\mathscr{N}}

\theoremstyle{remark}	\newtheorem{theorem}{Theorem}
\theoremstyle{remark}	\newtheorem{lemma}[theorem]{Lemma}
\theoremstyle{remark}	
\theoremstyle{remark}	
\theoremstyle{remark} \newtheorem{definition}{Definition}
\theoremstyle{remark} \newtheorem{remark}{Remark}
\theoremstyle{remark}

\begin{document}

\title{Secure Communication with Unreliable Entanglement Assistance}

\author{
    \IEEEauthorblockN{Meir Lederman and
    Uzi Pereg} \\
		\vspace{0.25cm}
    \IEEEauthorblockA{\normalsize
		\IEEEauthorrefmark{1}Faculty of Electrical and Computer Engineering, Technion \\
		\IEEEauthorrefmark{2}Helen Diller Quantum Center, Technion \\
    Email: {\tt 
 meirlederman@campus.technion.ac.il,     uzipereg@technion.ac.il}
}} 

\maketitle

\begin{abstract}
Secure communication is considered with unreliable entanglement assistance,
where the adversary may intercept the legitimate receiver's entanglement resource before communication takes place.
The communication setting of unreliable assistance, %
without security aspects, 
was originally  motivated by the extreme photon loss in practical communication systems. %
 The operational principle %
is to adapt the transmission rate
 to the availability of entanglement assistance, %
 without resorting to feedback and repetition. %
 Here, we require secrecy as well.
 An 
 achievable secrecy rate region is derived for general quantum wiretap channels, and a multi-letter secrecy capacity formula for the special class of degraded  channels. 
\end{abstract}


\section{Introduction}
Security poses a pivotal challenge in modern communication networks \cite{wang2020security, nguyen2021security,4529264, fettweis20216g,shamai2022information}.  %
Quantum key distribution (QKD) protocols aim to generate a shared key for subsequent message encryption
\cite{BennettBrassard:14p,Renner:08a,scarani2009security,pirandola2020advances}. 
Such protocols 
have been  implemented   experimentally 
 \cite{jouguet2013experimental, wang2015experimental, pugh2017airborne, liu2019experimental} and commercially \cite{qiu2014quantum, diamanti2016practical}.
Physical layer security complements the cryptographic key distribution approach, and leverages the inherent disturbance of the physical channel to ensure secure transmissions without relying on secret keys \cite{liang2013broadcast, schaefer2015secure}.
Wyner \cite{Wyner:75p} originally proposed this approach within the wiretap channel model.
A quantum wiretap channel is modeled as a linear map $\mathcal{N}_{A\to BE}$, %
where $A$ represents the system intended for transmission by the sender, Alice;  $B$ is the outcome obtained by a receiver,
Bob; and $E$ models what a malicious third-party, Eve, holds. 
The goal of secure communication
is for Alice to reliably transmit a  message to Bob while Eve gets negligible information
about the transmitted message.

Entanglement resources are useful in many quantum information applications, including   %
physical-layer security 
\cite{YLLYCZRCLL:20p,PeregDeppeBoche:21p,ZlotnickBashPereg:23c} and  network communication protocols
\cite{hsieh2008entanglement,DupuisHaydenLi:10p, pereg2023multiple}. 
In point-to-point communication, granting the transmitter and  receiver access to previously shared entanglement resources can significantly increase   transmission rate \cite{bennett1999entanglement, BennettShorSmolin:02p,QiSharmaWilde:18p, notzel2020entanglement, HaoShiLiShapiroZhuangZhang:21p}.
Unfortunately, entanglement is a fragile resource %
\cite{JaegerSergienko:14p, campbell2008measurement}.
In order to generate entanglement assistance in optical communication, the transmitter first prepares an entangled pair of particles locally, and then sends  %
one of them to the receiver \cite{YCLRLZCLLD:17p}. The procedure is not always successful in practice, as photons are easily  lost to the environment before reaching the  destination receiver \cite{czerwinski2022statistical, bergmann2016quantum}.

Current implementations typically involve an automatic repeat request (ARQ) mechanism. That is, the system incorporates the use of a back channel in order to notify the transmitter about the success or failure to establish entanglement between the transmitter and the receiver \cite{van2020extending}. 
In the case of a failure, the protocol is to be repeated. The ARQ approach has clear disadvantages from a system design perspective, as it may introduce a delay, further degradation of the entanglement resources, and even system collapse. 
Ensuring resilience and reliability is essential for trustworthiness and  critical for  developing of future communication networks \cite{FB22}.

Communication with unreliable entanglement assistance was recently introduced in \cite{pereg2023communication} as a setup where a back channel and repetition are not required.
Instead, the transmission rate is adapted to the availability of entanglement assistance. %
 The principle of operation of this model, as well as ours, ensures reliability by design for communication aided by entanglement resources (see further discussion in \cite{pereg2023communication-1b}). 
Uncertain cooperation was initially introduced in classical multi-user information theory  \cite{Steinberg:14c,HuleihelSteinberg:17p}. The idea is motivated by the engineering aspects of modern communication networks. In a dynamic ad-hoc communication setup, resources, such as energy, bandwidth, and time slots, are not guaranteed a priori. Their availability depends on parameters beyond the network designer's control, such as the battery status of intermediate users, weather conditions, or network peers'  willingness  to assist. Therefore, a likely scenario is that the users are aware of the possibility that cooperation will take place, but it cannot be assured before transmission begins. The classical models of unreliable cooperation mainly focus on dynamic resources in multi-user settings \cite{HuleihelSteinberg:16c, ItzhakSteinberg:21p, ItzhakSteinberg:17c, PeregSteinberg:21c4}. Whereas, the quantum model involves a point-to-point quantum channel and  unreliable  static resources of correlation.
Nonetheless, the seminal papers \cite{pereg2023communication,pereg2023communication-1b} did not consider security aspects.

The secrecy capacity of a quantum wiretap channel
has been investigated in various settings \cite{datta2010universal, smith2008private, dupuis2013locking, guha2014quantum, fawzi2013low, boche2017classical, tikku2020non, anshu2020secure, wiese2022mosaics, boche2022mosaics}. 
Cai  \cite{cai2004quantum} and Devetak \cite{devetak2005private} established 
a regularized multi-letter formula for the secrecy %
capacity of a quantum channel without entanglement assistance.  %
While a regularized characterization can be difficult to compute, it yields computable lower bounds, as well as a single-letter capacity formula for the special class of degradable channels \cite{devetak2005private, DevetakShor:05p}. 
Qi et al. \cite{QiSharmaWilde:18p} considered secure communication with entanglement assistance.
In principle, if the transmitter and receiver  share perfect entanglement beforehand, it can be utilized to generate a joint key, hence key distribution is not needed.
 Their model, however, does not allow Alice and Bob to generate a secret key in this manner, as Qi et al. \cite{QiSharmaWilde:18p} assume that the eavesdropper has access to the legitimate receiver's entanglement resource. %
While the assumption that both Bob and Eve can measure the same system may seem to contradict the no-cloning theorem, the scenario we describe below provides operational meaning to their assumption.

In this paper, we consider secure communication %
over a quantum wiretap channel with unreliable entanglement assistance.
Our model can be viewed as the combination of the two settings, i.e., communication with unreliable entanglement assistance \cite{pereg2023communication} and secure communication with (reliable) entanglement assistance \cite{QiSharmaWilde:18p}.
Before communication begins, the legitimate parties try to generate entanglement assistance. To this end, Alice prepares an entangled pair of particles locally and transmits one of them. While the particle travels from the transmitter, Eve tries to steal it.
In the optimistic case, Alice and Bob generate entanglement  successfully   prior to the transmission of information. Hence, Bob can decode the information while using the entangled resource, which is not available to Eve.
However, in the pessimistic case, Eve intercepts the entanglement resource, so Bob must decode without it.  In other words, Alice and \emph{Eve} share the entanglement, instead of Bob.
Nonetheless, despite the entangled resource being in Eve's possession, secrecy needs to be maintained.

Alice %
encodes two messages at rates $R$ and $R'$, while she does not know whether Bob or Eve hold the entanglement resource. 
Whereas, Bob knows whether entanglement assistance is available.
 In practice, this is realized through heralded entanglement generation \cite[Remark 2]{pereg2023communication}.
If Eve has stolen the entangled resource, then Bob decodes the first message alone; hence, the transmission rate is $R$.
Whereas, given entanglement assistance, Bob decodes both messages, hence the overall rate is $R+R'$.
The rate $R$ is thus associated with information that is guaranteed to be sent, while $R'$ corresponds to the excess information that entanglement assistance provides.
In this manner, we adapt the transmission rate to the availability of entanglement assistance, while communication does not collapse when the assisting resource is absent.

We establish an %
achievable rate region for secure communication with unreliable entanglement assistance and   a %
multi-letter formula of the secrecy capacity region for the class of degraded wiretap channels. %
The analysis  modifies the quantum superposition coding (SPC) scheme  in \cite{pereg2023communication}, by inserting local randomness elements that are used in the encoding of both messages in order to confuse Eve. %
Note that as opposed to traditional privacy models, in our model Eve is not completely passive, as she may capture the entanglement resource. This fundamental difference adds another layer of depth to our model and its characterization. %
While the setting resembles layered secrecy broadcast models \cite{zou2015broadcast, pereg2021key}, the analysis is much more involved, and the formulas have a different form. In particular, the bound on the guaranteed rate includes the entanglement resource along with Eve's system $E$.

The paper is organized as follows. In Section \ref{section_definitions}, we provide definitions and model description. %
In Section~\ref{Section:Previous_Work}, we give a brief review of related work. Our results are presented in Section~\ref{Section:Results}. The 
achievability proof for the inner bound is given in Section~\ref{Proof_Achievability}, and the proof for the regularized characterization for degraded wiretap channels in 
Section~\ref{Proof_Converse}. Section \ref{Section:Discussion} is dedicated to summary and discussion.

\section{Definitions}
\label{section_definitions}
\subsection{Basic Definitions}
We use the following notation conventions: Calligraphic letters  $\mathcal{X}, \mathcal{Y}, \mathcal{Z}, \ldots$ denote finite sets.
Uppercase letters $X, Y, Z, \ldots$ represent random variables, while the
lowercase letters $x, y, z, \ldots$ stand for their 
values. We use $x^j = (x_1, x_2, \ldots, x_j)$ for a sequence of letters from the alphabet $\mathcal{X}$, and $[i:j]$ denotes the index set $\{i, i+1, \ldots, j\}$ where $j>i$.

A quantum state  is described by a density operator $\rho$ on a Hilbert space $\mathcal{H}$.
We denote the set of all density operators by $\mathscr{S}(\mathcal{H}_A)$.
The von Neumann entropy is defined as $H(\rho)\equiv -\trace[\rho \log(\rho)]$. Given a bipartite state $\rho_{AB}\in\mathscr{S}(\mathcal{H}_A\otimes \mathcal{H}_B)$,
the quantum mutual information is 
$%
I(A;B)_\rho = H(\rho_{A})+H(\rho_{B})-H(\rho_{AB}) %
$. %
 The conditional quantum entropy is defined by $H(A|B)_\rho=H(\rho_{AB})-H(\rho_{B})$, and the quantum conditional mutual information is defined accordingly. %

A quantum channel $\mathcal{L}_{A\to B} : \mathscr{S}(\mathcal{H}_{A})\to \mathscr{S}(\mathcal{H}_{B})$ is a linear completely-positive and trace-preserving (CPTP) map. 

\subsection{Wiretap Channel}
\label{subsec:Qchannel}
A quantum wiretap channel $\mathcal{N}_{A\to BE}: \mathscr{S}(\mathcal{H}_{A})\to \mathscr{S}(\mathcal{H}_{B}\otimes \mathcal{H}_{E})$ maps a state at the sender's system to a joint state of the legitimate receiver  and eavesdropper's systems. The sender, receiver, and eavesdropper are often referred to as Alice, Bob and Eve, respectively.

We denote the marginal channel, from Alice to Bob, by $\mathcal{L}_{A\to B}$, and the marginal
 from Alice to Eve by
$\overline{\mathcal{L}}_{A\to E}$.
The marginal channels are also referred to as 
the main channel and the eavesdropper's channel, respectively.
The quantum wiretap channel $\mathcal{N}_{A\to BE}$ is called \emph{degraded} if there exists a degrading channel $\mathcal{P}_{B\to E}$ such that
\begin{align}
\overline{\mathcal{L}}_{A\to E}= \mathcal{P}_{B\to E}\circ \mathcal{L}_{A\to B}
\end{align}

We assume that the channel is memoryless, i.e., if Alice sends a sequence of input systems $A^n\equiv (A_1,\ldots, A_n)$, then the channel input $\rho_{A^n}$ undergoes the tensor-product mapping  $\mathcal{N}^{\otimes n}_{A\to BE}$.

\subsection{Coding with Unreliable Assistance}
\label{subsec:Mcoding}

\begin{figure}[tbp]
\centering
        \includegraphics[height=5cm, trim={3.5cm 10cm 0 10cm}, clip]{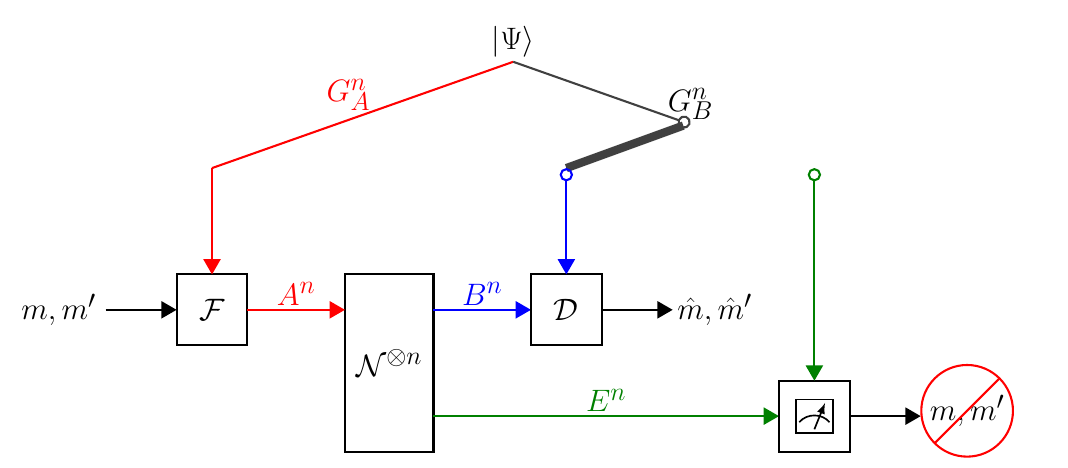} %
\\ \vspace{0.45cm}
(a)
\\ \vspace{1cm}
        \includegraphics[height=5cm, trim={3.5cm 10cm 0 10cm}, clip]{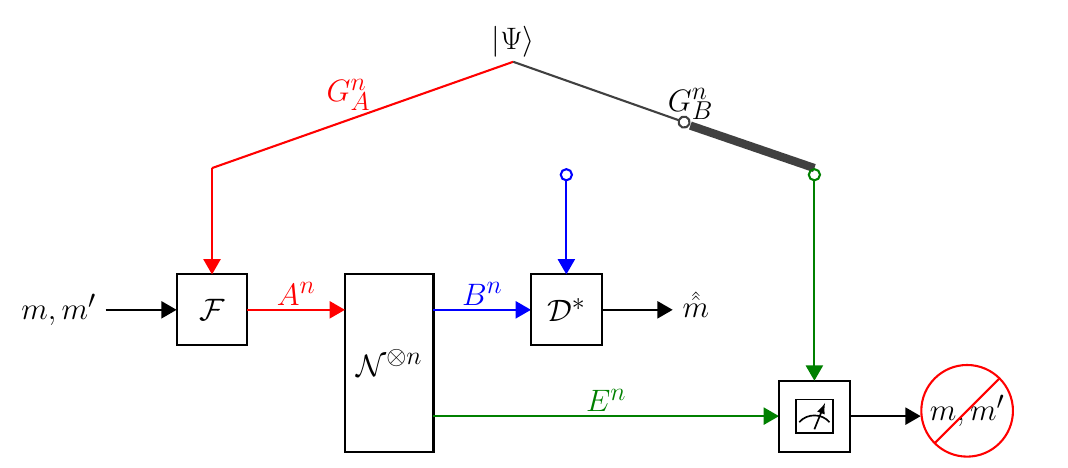} %
\\ \vspace{0.45cm}
(b)
   \vspace{0.85cm}  
    \caption{Illustration of secure  communication with unreliable entanglement assistance in the presence of an eavesdropper that may steal the entanglement resource. The figures show an imaginary switch that decides whether Eve will intercept the entanglement assistance or not. 
    There are two scenarios: (a) "left mode": Bob receives the entanglement resource and performs a measurement to decode both the guaranteed information and the excess information. (b) "right mode": Eve has intercepted the entanglement resource. Bob decodes the guaranteed information alone. The information needs to be secret in both scenarios.
    }
    \label{figure_switch}
\end{figure}

Before communication begins, the legitimate parties try to generate entanglement assistance. To this end, Alice prepares an entangled pair of particles locally, and transmits one of them. While the particle travels from the transmitter, Eve tries to steal it. 
In the optimistic case, Eve fails. Hence, Alice and Bob have pre-shared entanglement resources, $G_A^{n}$ and $G_B^{n}$, respectively.
That is,
entanglement is successfully generated prior to the transmission of information, and Bob can decode while using his access to $G_B^{n}$  (see Figure~\ref{figure_switch}(a)).
Otherwise, in the pessimistic case, Eve intercepts the entanglement resource $G_B^{n}$, in which case, Bob must decode without it.  In other words, the entanglement is shared between Alice and Eve, instead of Bob (see Figure~\ref{figure_switch}(b)).
Nonetheless, despite the entangled resource being stolen by Eve, secrecy needs to be maintained.

In the communication phase, Alice remains unaware of whether Bob or Eve possesses the entanglement resource. 
However, based on the common use of heralded entanglement generation in practical systems \cite{barz2010heralded}, we posit the assumption that both Bob and Eve are aware of the recipient of $G_B^{n}$ (see also \cite[Remark 2]{pereg2023communication}).

\begin{definition} 
\label{Definition:Capacity}
A $(2^{nR},2^{nR'},n)$  secrecy code with unreliable entanglement assistance consists of the following:
\begin{itemize}
\item
Two message sets $[1:2^{nR}]$ and $[1:2^{nR'}]$, where $2^{nR}$ and $2^{nR'}$ are assumed to be integers, 
\item
a pure entangled state $\Psi_{G^{n}_A, G^{n}_B}$, 
\item
a collection of encoding maps $\mathcal{F}_{G^{n}_A\xrightarrow{}A^n}^{(m,m')}: \mathscr{S}(\mathcal{H}_{G^{n}_A}) \to \mathscr{S}(\mathcal{H}_A^{\otimes n})$ for $m\in [1:2^{nR}]$ and $m'\in [1:2^{nR'}]$,  and 

\item
two decoding POVMs $\mathcal{D}_{B^nG^{n}_B}= \{D_{m,m'}\}$ and $\mathcal{D}^{*}_{B^n}= \{D^{*}_{m}\}$.
\end{itemize}
We denote the code by $(\Psi,\mathcal{F},\mathcal{D},\mathcal{D}^*)$.
\end{definition}

The communication scheme is depicted in Figure \ref{figure_switch}. Alice selects two messages, $m$ and $m'$, uniformly at random from  $[1:2^{nR}]$ and $[1:2^{nR'}]$, respectively. In addition, Alice  holds the resource $G^{n}_A$. %
She prepares the input state by applying the encoding map,
\begin{align}
\rho_{A^nG^{n}_{B}}^{m,m'} = \left(\mathcal{F}_{G^{n}_A\xrightarrow{}A^n}^{(m,m')}\otimes\mathrm{id}\right)
\left(\Psi_{G^{n}_AG^{n}_B}\right)
\label{encoding_messages}
\end{align}
and transmits $A^n$ through $n$ uses of the quantum wiretap channel. The channel output of Bob and Eve is 
\begin{align}
\rho_{B^n E^n G^{n}_B}^{m,m'} =(\mathcal{N}_{A\to B E}^{\otimes n}\otimes \identity)(\rho_{A^nG^{n}_B}^{m,m'}) \,.
\end{align}
Bob receives $B^n$.
As opposed to Alice, both Bob and Eve know whether they hold $G_B^{n}$ or not (thanks to heralded entanglement 
generation).
Depending on the availability of the entanglement assistance, he decides whether to decode both messages or only the guaranteed information. Given entanglement assistance, Bob has access to $G^{n}_B$, in which case he performs a measurement using the POVM $\mathcal{D}_{B^nG^{n}_B}= \{D_{m,m'}\}$ to recover both messages.  If Eve has sabotaged the entanglement assistance, then Bob recovers the message $m$ alone, using the POVM $\mathcal{D}^*_{B^n}= \{D^*_{m}\}$.

Bob is required to decode correctly the guaranteed information $m$, whether the entanglement resource is stolen or not. Whereas, Bob need only decode the excess information $m'$ if the entanglement was not stolen and thus available to him. In both cases, Alice and Bob need to maintain full secrecy from Eve.

\subsubsection{Correct Decoding}
Since there are two scenarios in our setting, we also have two error criteria. In the presence of entanglement assistance, Bob decodes with $\mathcal{D}_{B^nG^{n}_B}= \{D_{m,m'}\}$. Hence, the average probability of error,

\begin{align}
P^{(n)}_{e}(\Psi,\mathcal{F},\mathcal{D}) 
&= 1 - \frac{1}{2^{n(R+R')}}\sum_{m=1}^{2^{nR}}\sum_{m'=1}^{2^{nR'}}\trace\left\{D_{m,m'} \,
\rho_{B^n G_B^n}^{m,m'}
\right\}
\nonumber\\
&= 1 - \frac{1}{2^{n(R+R')}}\sum_{m=1}^{2^{nR}}\sum_{m'=1}^{2^{nR'}}\trace\left\{D_{m,m'}(\mathcal{L}_{A\xrightarrow{}B}^{\otimes n}\otimes \mathrm{id})(\mathcal{F}_{G^{n}_A\xrightarrow{}A^n}^{(m,m')}\otimes  \mathrm{id})(\Psi_{G^{n}_A,G^{n}_B})\right\}
\end{align}
In the absence of entanglement assistance, the average error probability is:

\begin{align}
P^{*(n)}_{e}(\Psi,\mathcal{F},\mathcal{D}^*)
&= 1 - \frac{1}{2^{n(R+R')}}\sum_{m=1}^{2^{nR}}\sum_{m'=1}^{2^{nR'}}\trace\left\{D_{m}^* \,
\rho_{B^n}^{m,m'}
\right\}
\nonumber\\
&= 1 - \frac{1}{2^{n(R+R')}}\sum_{m=1}^{2^{nR}}\sum_{m'=1}^{2^{nR'}}\trace\left\{D^*_{m}(\mathcal{L}_{A\xrightarrow{}B}^{\otimes n}\circ\mathcal{F}_{G^{n}_A\xrightarrow{}A^n}^{(m,m')})(\Psi_{G^{n}_A})\right\}
\end{align}

\subsubsection{Secrecy}
Confidentiality requires that the correlation between Eve's output and the messages is negligible. Formally, 
the leakage is measured by
\begin{align}
\ell^{(n)}(\Psi,\mathcal{F})= 
I(M,M';E^n, G^{n}_B)_\rho \,,
\label{leakage_definiton}
\end{align}
where
\begin{align}
\rho_{MM'B^{n}E^{n}G_{B}}&= 
\frac{1}{2^{n(R+R')}}\sum_{m=1}^{2^{nR}}\sum_{m'=1}^{2^{nR'}}
\ketbra{m}_{M}\otimes \ketbra{m'}_{M'} \otimes \rho^{m,m'}_{B^{n}E^{n}G_{B}}\,, 
\\
\rho_{B^{n}E^{n}G_{B}}&= (\mathcal{N}_{A\to B E}^{\otimes n}\otimes \mathrm{id})(\rho_{A^n}\otimes \Psi_{G_{B}}) \
\end{align}
where the classical registers $M$ and $M'$ are uniformly distributed over
the message sets $[1:2^{nR}]$ and $[1:2^{nR'}]$, respectively.
This criterion is referred to as strong secrecy.
Notice that we include the entangled resource $G^{n}_B$ in the leakage criterion due to the pessimistic case where Eve obtains $G^{n}_B$ instead of Bob.

\subsubsection{Achievable rate pairs}
A $(2^{nR}, 2^{nR'}, n, \epsilon, \delta)$ secrecy code with unreliable entanglement assistance satisfies that the error probabilities  are bounded by $\epsilon$, and the leakage by $\delta$. That is, 
\begin{align}
&\max\left(P^{(n)}_{e}(\Psi,\mathcal{F} \,,\; \mathcal{D}),P^{*(n)}_{e}(\Psi,\mathcal{F},\mathcal{D}^*)\right)\leq\epsilon \,, 
\label{error_demand}
\intertext{and additionally,} 
&\ell^{(n)}(\Psi,\mathcal{F})\leq\delta \,.
\label{leakage_demand} 
\end{align}

A secrecy rate pair $(R,R')$ is called achievable if for every $\epsilon,\delta> 0$ and sufficiently large $n$, there exists a $(2^{nR}, 2^{nR'}, n, \epsilon,\delta)$ secrecy code with unreliable entanglement assistance.
The secrecy capacity region with unreliable entanglement assistance,  denoted by $\mathcal{C}_{\text{S-EA*}}(\mathcal{N})$, is the closure of the set of all achievable rate pairs.

\begin{remark}
\label{Remark:Shared_Key}
A straightforward method to leverage entanglement assistance is to generate a shared 
key, and then encode the information using the one-time pad protocol. However, this strategy poses a security risk in our case. If Eve  intercepts the entanglement resource, then she will get a hold of Alice's  key, resulting in a failure and a breach of security. 
\end{remark}

\begin{remark}
\label{Remark:Superdense_Coding}
 Eve's interception has severe consequences on entanglement-assisted communication. 
For example, 
suppose that Alice uses the super dense coding protocol to encode two classical bits, and then transmits her qubit  via a quantum erasure channel.
Consider the event that Bob receives an erasure, hence Eve receives the transmitted qubit.
Nevertheless, without the entanglement resource, there is no leakage, because each qubit by itself has no correlation with Alice's messages. On the other hand, if Eve has both qubits, then she can use the super dense decoder in order to recover Alice's bits.
\end{remark}

\begin{remark}
\label{Remark:Hard_Decicion}
Our model considers two extreme scenarios, i.e.,
 the entanglement resources are either entirely available to Bob or not at all. %
In digital communications, this strategy aligns with a \emph{hard decision} approach \cite{Proakis:01b}. Indeed, the decoder in our setting makes a hard decision on whether the  entanglement resources are viable. 
This approach fundamentally differs from noisy entanglement models that ensure reliability with respect to the average state 
 \cite{ZhuangZhuShor:17p}.

\end{remark}

\begin{remark}
\label{Remark:Guaranteed_Message_Correlation}
We observe that guaranteed information could have correlation with the receiver's entanglement resource.
Indeed, the guaranteed information $m$ needs to be encoded in such a manner that Bob could recover it even in the absence of the entanglement resource, see Figure~\ref{figure_switch}(b). 
Nevertheless,   Alice encodes \emph{her} resource $G_{A}^{n}$ using an encoding map that depends on both $m$ and $m'$ (see \eqref{encoding_messages}).
 As a result, the encoding operation may induce correlation between the guaranteed information $m$ and the entangled resource $G_{B}^{n}$.
 We will see the consequences of this observation on the rate region formula in Section~\ref{Section:Results} below, see Remarks \ref{Remark:Strange_Form} and \ref{Remark:Formula_Correlation_XG2}.
    \label{remark_encoding_G_B}
\end{remark}

\begin{figure*}[tb]	
\begin{center}
\hspace{-2cm}
\begin{tabular}{l|cc}
&$\;$   Unsecure $\;$ 
&       Secure $\;$    
\\	[0.2cm]   
\hline 
\\			[-0.2cm]
No assistance				
&	$C_0(\mathcal{L})$
&   $C_{\text{S}}(\channel)$ 	
\\[0.3cm] 
Reliable Assistance					
&	$C_{\text{EA}}(\mathcal{L})$
&   $C_{\text{S-EA}}(\mathcal{N})$ 						
\\[0.3cm]
Unreliable Assistance					
&	 $\mathcal{C}_{\text{EA*}}(\mathcal{L})$
&    $\mathcal{C}_{\text{S-EA*}}(\mathcal{N})$ 		
\end{tabular}
\end{center}
  \caption{Notation of channel capacities with and without secrecy, and with different levels of entanglement assistance. The first column corresponds to communication without a secrecy requirement, and the second column comprises  secrecy capacities. %
	}
\label{table:LcapacityNotation}
\end{figure*}

\section{Previous Work}
\label{Section:Previous_Work}
We provide a brief review of known results  with and without secrecy, and with different levels of entanglement assistance. We denote the corresponding capacities
as in Figure~\ref{table:LcapacityNotation}. 
Specifically, we denote the non-secure capacities,  without entanglement assistance, with reliable entanglement assistance, and with unreliable entanglement assistance,  by $C_0(\mathcal{L})$, $C_{\text{EA}}(\mathcal{N})$, and $\mathcal{C}_{\text{EA*}}(\mathcal{N})$, respectively.
Similarly, we denote the secrecy capacities by
 $C_{\text{S}}(\mathcal{N})$,  $C_{\text{S-EA}}(\mathcal{N})$, and $\mathcal{C}_{\text{S-EA*}}(\mathcal{L})$, respectively.

\subsection{Unsecure Communication}
At first, we consider communication %
over a quantum channel $\mathcal{L}_{A\to B}$, without a secrecy requirement.
We review the capacity results without assistance, with reliable entanglement assistance, and with unreliable entanglement assistance.

\subsubsection{No Assistance}
Suppose that Alice and Bob do not share entanglement a priori.
The Holevo information of the channel 
is defined as
\begin{align}
    \chi(\mathcal{L}) &\equiv \max_{p_{X}(x), \ket{\psi_{A}^x}} I(X;B)_{\omega} \,,
    \intertext{where the maximization is over the ensemble of quantum input states, and}
    \omega_{XB} &\equiv \sum_{x\in\mathcal{X}} p_{X}(x)\ketbra{x} \otimes \mathcal{L}(\ketbra{\psi_{A}^{x}}) %
\label{Equation:Holveo_Information}
\end{align}
with $\abs{\mathcal{X}} \leq \mathrm{dim}(\mathcal{H}_{A})^2$.
\begin{theorem}[see \cite{Holevo:98p,SchumacherWestmoreland:97p}] 
\label{Theorem:Without_Assistance}
The  capacity of a quantum channel $\mathcal{L}_{A\to B}$ without secrecy and without assistance satisfies  
\begin{align}
    C(\mathcal{L}) = \lim_{n\to\infty}\frac{1}{n}\chi(\mathcal{N}^{\otimes n})
\end{align}
\end{theorem}

\subsubsection{Reliable Assistance}
Consider entanglement-assisted communication, where it is assumed that the entanglement assistance is reliable with certainty.
Let
\begin{align}
    I_{\text{EA}}(\mathcal{L}) &= \max_{\ket{\phi_{GA}}} I(G ;B)_\omega \,,
    \intertext{where the maximum is over all bipartite states $\ket{\phi_{GA}}$, and
    }
    \omega_{G B} &= (\text{id}\otimes\mathcal{L})(\ketbra{\phi_{G A}}) \,,\;
\label{Equation:Channel_Mutual_Information}
\end{align}
with $\mathrm{dim}(\mathcal{H}_{G }) \leq \mathrm{dim}(\mathcal{H}_{A})$.
The system $G$ in \eqref{Equation:Channel_Mutual_Information} can  be interpreted as Bob's entanglement resource.

\begin{theorem}[see \cite{BennettShorSmolin:99p}] 
\label{Theorem:Reliable_Assistance}
The  capacity of a quantum channel $\mathcal{L}_{A\to B}$ without secrecy and with (reliable) entanglement assistance  is given by 
\begin{align}
    C_{\text{EA}}(\mathcal{N}) &= I_{\text{EA}}(\mathcal{L})
\end{align}
\end{theorem}

\subsubsection{Unreliable Assistance}
We move to communication with \emph{unreliable} entanglement assistance, as presented in Subsection~\ref{subsec:Mcoding}, but without a secrecy requirement. Recall that Alice sends two messages, a guaranteed message at rate $R$ and excess message at rate $R'$, hence the performance is characterized by rate regions. %

Define 
\begin{align}
\mathcal{R}_\text{EA*}(\mathcal{L})
&=
\bigcup_{ p_X, \varphi_{G_{1}G_{2}}, \mathcal{F}^{(x)} } 
\left\{ \begin{array}{rl}
  (R,R') \,:\;
    R   \leq    &    I(X;B)_\omega \\
    R'  \leq    &    I(G_{2};B|X)_\omega \\
	\end{array}
\right\} \,,
\intertext{where}
    \omega_{XG_{2}A}&=\sum_{x\in\mathcal{X}}p_{X}(x)\ketbra{x}\otimes(\text{id}\otimes\mathcal{F}^{(x)}_{G_{1}\to A})(\varphi_{G_{2}G_{1}})
\intertext{and}
\omega_{XG_{2}B}&=(\text{id}\otimes\mathcal{L}_{A\to B})(\omega_{XG_{2}A}) \,.
\end{align}

\begin{theorem}[see \cite{pereg2023communication}] 
\label{Theorem:Unreliable_Assistance}
The  capacity region of a quantum channel $\mathcal{L}_{A\to B}$ with \emph{unreliable} entanglement assistance satisfies
\begin{align}
    \mathcal{C}_\text{EA*}(\mathcal{L}) = 
    \bigcup_{n=1}^\infty
    \frac{1}{n}\mathcal{R}_\text{EA*}(\mathcal{L}^{\otimes n}) \,.
\end{align}
\end{theorem}

\subsection{Secure Communication}
Consider a quantum wiretap channel $\mathcal{N}_{A\to BE}$.
Here, we review the fundamental capacity results without assistance and with reliable entanglement assistance, while secure communication with unreliable entanglement assistance will be addressed in the results section.

\subsubsection{No Assistance}
Suppose that Alice and Bob do not share entanglement a priori.
The private information of the quantum wiretap channel is defined by
\begin{align}
    I_{\text{S}}(\mathcal{N}) &\equiv \max_{p_{X}(x), \omega_{A}^x} [I(X;B)_{\omega}-I(X;E)_{\omega}] \,, 
    \label{definition_private_information}
\intertext{where the maximization is as in \eqref{Equation:Holveo_Information}, and}    
    \omega_{XBE} &\equiv 
    \sum_{x\in\mathcal{X}}p_X(x)\ketbra{x}\otimes \mathcal{N}_{A\to BE}(\omega_{A}^x) \,,
\end{align}
with $\abs{\mathcal{X}} \leq \mathrm{dim}(\mathcal{H}_{A})^2+1$.
\begin{theorem}[see \cite{cai2004quantum,devetak2005private}] 
\label{Theorem:Secrecy_Without_Assistance}
The secrecy capacity of a quantum wiretap channel $\mathcal{N}_{A\to BE}$ without assistance is given by 
\begin{align}
    C_{\text{S}}(\mathcal{N}) = \lim_{k\to\infty}\frac{1}{k}I_{\text{S}}(\mathcal{N}^{\otimes k}) 
    \label{capacity_wiretap_channel}
\end{align}

\end{theorem}
A single-letter formula for the secrecy capacity remains an open problem for a  general quantum wiretap channel. 
The secrecy capacity and the private information are both known to be super additive
\cite{ElkoussStrelchuk:15p,li2009private}.

\subsubsection{Reliable Assistance}
Consider entanglement-assisted communication under a secrecy requirement, where it is assumed that the entanglement assistance is reliable with certainty.
Define
\begin{align}
    I_{\text{S-EA}}(\mathcal{N}) = \max_{\varphi_{AG}}[I(G;B)_\omega - I(G;E)_\omega] \,, 
    \intertext{where the maximum is over all bipartite states $\varphi_{AG}$, and}
    \omega_{GBE} \equiv (\text{id}\otimes\mathcal{N}_{A\to BE})(\varphi_{GA})  \,.
\end{align}
\begin{theorem} [see \cite{QiSharmaWilde:18p}]
\label{Theorem:Secrecy_Reliable_Assistance}
The secrecy capacity of a quantum wiretap channel $\mathcal{N}_{A\to BE}$ with (reliable) entanglement assistance  is bounded by %
\begin{align}
    C_{\text{S-EA}}(\mathcal{N}) \geq   I_{\text{S-EA}}(\mathcal{N}) \,.
\end{align}
Furthermore, if the channel is degraded, then
\begin{align}
    \mathcal{C}_{\text{S-EA}}(\mathcal{N}) =   I_{\text{S-EA}}(\mathcal{N}) \,.
\end{align}
\end{theorem}

A single-letter formula for the entanglement-assisted secrecy capacities is an open problem as well.

\section{Results}
\label{Section:Results}
We consider  secure communication with unreliable entanglement assistance. Recall that  Alice does not know whether the entanglement resource has reached Bob's location, hence  she encodes two messages at rates $R$ and $R'$ (see Subsection~\ref{subsec:Mcoding}). %
If entanglement assistance is available to Bob, he recovers both messages. Yet, if Eve has stolen the resource, then recovers the first message alone.
Nonetheless, we require  the information to be secret from Eve in both scenarios.

\subsection{Inner Bound}
First, we establish an achievable secrecy rate region. %
Let $\mathcal{N}_{A\to BE}$ be a quantum wiretap channel. Define
\begin{align}
\mathcal{R}_\text{S-EA*}(\mathcal{N})
&\equiv
\bigcup_{ p_X, \varphi_{G_{1}G_{2}}, \mathcal{F}^{(x)} } 
\left\{ \begin{array}{rl}
  (R,R') \,:\;
    R   \leq    &    [I(X;B)_\omega - I(X;EG_{2})_\omega]_+  \\
    R'  \leq    &    [I(G_{2};B|X)_\omega - I(G_{2};E|X)_\omega]_+ \\
	\end{array}
\right\} \ 
\label{R_S-EA*}
\end{align}
where
\begin{align}
    \omega_{XG_{2}A}&\equiv\sum_{x\in\mathcal{X}}p_X(x)\ketbra{x}\otimes(\text{id}\otimes\mathcal{F}^{(x)}_{G_{1}\to A})(\varphi_{G_{2}G_{1}}) \,,
    \label{Equation:omegaXG2A}
\\
    \omega_{XG_{2}BE}&\equiv(\text{id}\otimes\text{id}\otimes\mathcal{N}_{A\to BE})(\omega_{XG_{2}A}) \,,
\end{align}
with $[x]_+\equiv \max(x,0)$.
Our main result is given in the theorem below. %
\begin{theorem}
\label{Theorem:Inner_Bound}
The region $\mathcal{R}_\text{S-EA*}(\mathcal{N})$ is an achievable secrecy rate region with unreliable entanglement assistance. That is,
the  secrecy capacity region with unreliable entanglement assistance is bounded by
\begin{align}
\mathcal{C}_{\text{S-EA*}}(\mathcal{N}) \supseteq
\mathcal{R}_\text{S-EA*}(\mathcal{N})
\end{align}
\end{theorem}

The proof of Theorem \ref{Theorem:Inner_Bound} is given in section \ref{Proof_Achievability}. 
We  modify the quantum superposition coding (SPC) scheme  in \cite{pereg2023communication} by inserting local randomness elements that are used in the encoding, one for each  message, in order to confuse Eve. In the analysis, we use the quantum covering lemma \cite{ahlswede2002strong} in a non-standard manner. 
\begin{remark}
\label{Remark:Strange_Form}
While the setting resembles layered secrecy broadcast models \cite{zou2015broadcast, pereg2021key}, the analysis is much more involved, and the formulas have a different form.
Specifically, instead of the mutual information term 
$I(X;E)_\omega$ in the private information formula, %
we now have $I(X;EG_2)_\omega$ that includes the receiver's entanglement resource, cf.
  \eqref{definition_private_information} and \eqref{R_S-EA*}.
\end{remark}

\begin{remark}
\label{Remark:Formula_Correlation_XG2}
Based on the model description,
 it may seem at a first glance as if $X$ should not be correlated with $G_2$, since the guaranteed information needs to be recovered in the absence of the entanglement resource.  
However, 
 as pointed out in Remark~\ref{remark_encoding_G_B}, %
   Alice's encoding %
   may induce correlation between the guaranteed information and the receiver's resource. %
Similarly, in the rate region formula, the application of the encoding channel $\mathcal{F}_{G_1\to A}^{(x)}$ could create correlation between $X$ and $G_2$ (see \eqref{Equation:omegaXG2A}).

\end{remark}

\begin{remark}
In the coding scheme, we specified that Bob applies one of two distinct POVMs, depending on who holds the entanglement resource --- Bob or Eve. If Eve has sabotaged his assistance, Bob performs $\mathcal{D}^{*}_{B^n}= \{D^{*}_{m}\}$ to decode  $m$ alone. Otherwise, if Bob has the assistance, he performs $\mathcal{D}_{B^nG^{n}_B}= \{D_{m,m'}\}$ to decode both $m$ and $m'$. Nonetheless, the quantum SPC scheme {\cite{pereg2023communication}} employs a sequential decoder.
On the first stage, Bob performs a measurement to obtain an estimate for the guaranteed message $m$. The gentle measurement lemma \cite{Pereg:21p, Winter:99p, OgawaNagaoka:07p} guarantees that there is no collapse after the first measurement, i.e.,  the output state  remains nearly unchanged. 
Then, Bob moves on to the second stage. In the presence of the entanglement resource, Bob performs a second measurement to estimate the excess message $m'$, and in the absence of his resource, he aborts.

\end{remark}

\subsection{Regularized Capacity Formula}
For the class of  degraded channels, we establish a multi-letter capacity formula.

\begin{theorem}
    Let $\mathcal{N}_{A\to BE}$ be a  degraded  quantum wiretap channel. The unreliable entanglement assisted secrecy capacity region satisfies
    \begin{align}
    \mathcal{C}_{\text{S-EA*}}(\mathcal{N}) = 
        \bigcup_{n=1}^\infty
        \frac{1}{n}\mathcal{R}_{\text{S-EA*}}(\mathcal{N}^{\otimes n}) 
    \end{align}
    \label{Theorem:Degraded_Channels}
\end{theorem}
The proof of Theorem \ref{Theorem:Degraded_Channels} is given in section \ref{Proof_Converse}.

\section{Proof of Theorem~\ref{Theorem:Inner_Bound}}
\label{Proof_Achievability}

Consider secure communication with unreliable entanglement assistance.
We show that every secrecy rate pair $(R,R')$ in the interior of $\mathcal{R}_{\text{S-EA*}}(\mathcal{N})$ is achievable.
Suppose  Alice wishes to send a pair of messages, $(m,m')\in [1:2^{nR}]\times [1:2^{nR'}]$. %
In the optimistic case, entanglement is successfully generated prior to the transmission of information, hence Bob can decode while using the entangled resource, which is not available to Eve.
However, in the pessimistic case, Eve intercepts the resource, in which case, Bob must decode without it.
The coding scheme modifies the quantum SPC construction from \cite{pereg2023communication}. 
Here, we insert local randomness elements, which will be denoted in the analysis as $k,k'$, and are used in the encoding of each message in order to confuse Eve.
Our secrecy analysis relies on the quantum covering lemma \cite{ahlswede2002strong}, as stated below.
The quantum covering lemma can be viewed as a direct consequence of quantum channel resolvability \cite[Appendix B]{tahmasbi2020toward}, \cite{hayashi2006quantum}.

\begin{lemma}
\label{quantum_covering_lemma}
let $\{p_{X}(x), \sigma_{x}\}_{x\in\mathcal{X}}$ be an ensemble, with a mean state $\sigma \equiv \sum_{x\in\mathcal{X}}p_{X}(x)\sigma_x$.
Furthermore, suppose that there is a code projector $\Pi$ and codeword projectors $\{\Pi_{x}\}_{x\in\mathcal{X}}$, that satisfy for every $\epsilon>0$:
\begin{align}
    \trace \{ \Pi \sigma_{x}\} &\geq 1-\epsilon
    \\
    \trace\{\Pi_{x} \sigma_{x}\} &\geq 1-\epsilon
    \\
    \trace\{\Pi\} &\leq \mathsf{D}
    \\
    \Pi_{x} \sigma_{x} \Pi_{x} &\leq \frac{1}{\mathsf{d}}\Pi_{x}
\end{align}
where $0<\mathsf{d}<\mathsf{D}$.
Consider a random codebook that $\mathscr{C}\equiv \{X(k)\}_{k\in\mathcal{K}}$ that consists of $\abs{\mathcal{K}}$ statistically independent codewords, each distributed  according to $p_{X}(x)$. %
Then,
\begin{align}
    \Pr{\norm{\frac{1}{\abs{\mathcal{K}}}\sum_{k\in\mathcal{K}}\sigma_{X(k)}-\sigma}_1 > \epsilon+4\sqrt{\epsilon}+24\sqrt[4]{\epsilon}} \leq 1- 2\mathsf{D}\exp{-\frac{\epsilon^3}{4\ln2}\frac{\abs{\mathcal{K}}\mathsf{d}}{\mathsf{D}}} 
\end{align}
\end{lemma}

Before we begin the achievability proof, we make the following observations. First, we note that pure states $\ket{\phi_{G_{1}G_{2}}}$ are sufficient to exhaust the union in the rate region formula in \eqref{R_S-EA*}.
Indeed, every quantum state $\varphi_{G_1 G_2}$ has a purification $\ket{\phi_{G_1 J_1 G_2}}$. Since $G_1$ is arbitrary, we can extend it and obtain the same characterization when $G_1$ is replaced by $(G_1,J_1)$ 
(see also \cite[App. B]{pereg2023communication}).
Therefore, we may assume without loss of generality that 
$\varphi_{G_1 G_2}$ is pure, i.e.,
$\varphi_{G_1 G_2}=\ketbra{\phi_{G_1 G_2}}$.

In addition, we observe that we can restrict the proof to  isometric encoding maps, $F^{(x)}_{G_1 \to A}$ for $x\in\mathcal{X}$, by similar arguments as in \cite{pereg2023communication}. 
To see this, %
consider using a collection of encoding channels,  $\mathcal{F}^{(x)}_{G_1 \to A'}$ for $x\in\mathcal{X}$, for transmission via  $\widehat{\mathcal{N}}_{A'\to BE}$. Every quantum channel $\mathcal{F}^{(x)}_{G_1 \to A'}$ has a Stinespring representation, with an isometry $F^{(x)}_{G_1 \to A'A_0}$. 
Since it is an encoding map, we may think of $A_0$ as Alice's ancilla. Then, let $A\equiv (A',A_0)$ be the augmented channel input. We are effectively coding over the channel $\mathcal{N}_{A\to BE}$, where
$\mathcal{N}_{A\to BE}(\rho_{A'A_0})=\widehat{\mathcal{N}}_{A'\to BE}(\trace_{A_0}(\rho_{A'A_0}))$, using the isometric map $F^{(x)}_{G_1\to A}$.
From this point, we will focus on the quantum wiretap channel $\mathcal{N}_{A\to BE}$ and use the isometric encoding map $F^{(x)}_{G_1\to A}$.

\subsection{Notation} 
We introduce the following notation.
For every $x\in {\mathcal{X}}$, 
consider the input state
\begin{align}
    \ket{\psi^{x}_{AG_2}} &= (F_{G_{1}\to A}^{(x)}\otimes\identity)\ket{\phi_{G_{1}G_{2}}}
    \,,
    \label{define_psi}
\intertext{which results in the output}
    \omega^{x}_{BEG_2} &= (\mathcal{N}_{A\to BE}\otimes\text{id})(\psi^{x}_{AG_2})
    \,.
    \label{define_omega}
\end{align}
Then, consider
a Schmidt decomposition, %
\begin{align}
\label{psi_schmidt_decom}
    \ket{\psi^x_{AG_2}} = \sum_{y\in\mathcal{Y}}\sqrt{p_{Y|X}(y|x)}\ket{\xi_{y|x}}\otimes\ket{\xi^{'}_{y|x}}
\end{align}
where $p_{Y|X}$ is a conditional probability distribution.
We will often use the notation $\ket{\psi^{x^n}} = \bigotimes_{i=1}^{n}\ket{\psi^{x_i}}$.

Next, let us define a unitary operator that will be useful in the definition of our encoder. Denote the Heisenberg-Weyl operators, on a qudit of dimension $d$, by 
\begin{align}
\Sigma(a,b)=\Sigma_X^a \Sigma_Z^b \,,\;\text{for
$a,b\in\{0,\ldots,d-1\}$} \,,
\end{align}
where
$\Sigma_X=\sum_{k=0}^{d-1}\ket{k+1 \mod d}\bra{k}$ and $\Sigma_Z=\sum_{k=0}^{d-1}\exp{\frac{2\pi ik}{d}}\ketbra{k}$.

Let $x^n\in\mathcal{X}^n$ be a given sequence.
For every conditional type $t$ on $\mathcal{Y}^n$ given $x^n$, we will apply an operator of the form
$(-1)^{c_t}\Sigma(a_t,b_t)$ %
for $a_t,b_t\in\{0,\ldots,d_t-1\}$ and $c_t\in\{0,1\}$,
where 
$d_t$ is the size of the corresponding conditional type 
class.
Then, define the  unitary 
\begin{align}
    U(\gamma) &= \bigoplus_{t}(-1)^{c_t} \Sigma(a_t,b_t)
\end{align}
corresponding to a vector
    $\gamma=\left( (a_t,b_t,c_t)_t \right)$,
where the direct sum is %
over all conditional types. Furthermore, let $\Gamma_{x^{n}}$ denote the set of all such vectors $\gamma$.

\subsection{Code Construction}
We now describe the construction of a secrecy code with unreliable entanglement assistance.
Let $\ket{\phi_{G_{1}G_{2}}}^{\otimes n}$ be the assistance that Alice and Bob would like to share.
We also let $R_0$ and $R_0'$ denote the rates of the Alice's local random elements, where $0<R_0<R$ and $0<R_0'<R'$. 

\vspace{0.1cm}
\subsubsection{Classical Codebook Generation} 
Select $2^{n(R+R_0)}$  sequences, 
\begin{align}
\big\{x^{n}(m,k) \big\}_{
\substack{m\in [1:2^{nR}],\\ k\in [1:2^{nR_0}]}}
\end{align}
independently at random, each i.i.d.  $\sim p_X$.
Then, for every $m$ and $k$, select $2^{n(R'+R_0')}$ conditionally independent sequences, 
\begin{align}
\big\{ \gamma(m',k'|x^{n}(m,k)\big\}_{\substack{m'\in [1:2^{nR'}],\\ k'\in [1:2^{nR'_0}]}}
\end{align}
at random, each uniformly distributed over $\Gamma_{x^{n}(m,k)}$.
The codebooks are publicly revealed, to Alice, Bob, and Eve.

\subsubsection{Encoding} Alice  chooses a  message pair $(m,m')$, which is assumed to be uniformly distributed.  %
To ensure secrecy, Alice further selects local randomness elements, $k$ and $k'$, chosen uniformly at random, from $ [1:2^{nR_0}]$ and $ [1:2^{nR_0'}]$, respectively.
To encode the first message $m$, she %
applies the encoding map %
\begin{align}
F_{G_1^n\to A^n}^{(x^{n})}=\bigotimes_{i=1}^n
F_{G_1\to A}^{(x_i)} \,,\;\text{with $x^n\equiv x^{n}(m,k)$} \,,
\end{align}
 on her share of the entangled state 
 $\ket{\phi_{G_{1}G_{2}}}^{\otimes n}$.
The resulting input state is 
 $\ket{\psi^{x^n}_{A^{n}G_2^{n}}}$ (see \eqref{define_psi}).

To encode the excess message  $m'$, 
she applies the unitary $U(\gamma)$, with 
$\gamma\equiv \gamma(m',k'|x^{n}(m,k))$.
This yield the following input state, %
\begin{align}
    \ket{\chi^{\gamma, x^{n}}_{A^nG_2^n}} = (U(\gamma)\otimes\identity)\ket{\psi^{x^{n}}_{A^{n}G_2^n}} \,.
\end{align}
Alice transmits $A^n$ through $n$ uses of the quantum wiretap channel $\mathcal{N}_{A\to BE}$. The output state  is
\begin{align}
    \rho^{\gamma,x^{n}}_{B^{n}E^{n}G_2^{n}} = (\mathcal{N}_{A\to BE}^{\otimes n}\otimes\text{id})(\chi^{\gamma,x^{n}}_{A^{n}G_2^{n}}) \,.
\end{align}

\subsubsection{Decoding} %
Bob has two decoding strategies. 
If Bob holds the entangled resource $G_2^n$, then he decodes both messages, $m$ and $m'$. However, if Eve has stolen $G_2^n$, then Bob decodes the message $m$ alone.
Specifically, Bob decodes in two steps. First, he performs a measurement, using a POVM $\{\Lambda_{m,k}\}$, which will be described later, in order to estimate the message $m$.
If he has access to the entanglement resource $G_2^n$, then he continues to decode the message $m'$ using a second POVM $\{\Upsilon_{m',k'}\}$, which will also be described later.

\subsection{Error Analysis}
We now analyze the probability for erroneous decoding by Bob, for the guaranteed message and the excess message. 
Let $\alpha>0$ be arbitrarily small.
First, we establish properties of the code above.
Using the schmidt decomposition in \eqref{psi_schmidt_decom},
we have
\begin{align}
    \ket{\psi^{x^n}_{A^nG_2^n}} &= \sum_{y^{n}\in\mathcal{Y}^{n}}\sqrt{P_{Y^{n}|x^n}(y^n|x^n)}\ket{\xi_{y^n|x^n}}\otimes\ket{\xi^{'}_{y^n|x^n}}  
 \end{align}
 for $x^n\in\mathcal{X}^n$.
Following similar arguments as in \cite{pereg2023communication},
we can also write this as
 \begin{align}
     \ket{\psi^{x^n}_{A^nG_2^n}} &= %
    \sum_{t}\sqrt{p(t|x^{n})}\ket{\Phi_{t}}
    \label{typical_subspaces_decom}
\end{align}
where the sum is over all conditional types on $\mathcal{Y}^n$ given $x^n$,
with $p(t|x^{n})$ denoting the probability that a random sequence $Y^n\sim p^n_{Y|X}(\cdot|x^n)$ belongs to the conditional type class of $t$, and 
$\ket{\Phi_{t}}=\sum_{y^{n}\in T_{n}(t|x^n)}\ket{\xi_{y^n|x^n}}\otimes\ket{\xi^{'}_{y^n|x^n}}$ %
is a maximally entangled state on the product of typical subspaces (see \cite[Eq. (71)]{pereg2023communication}).
Using the ricochet property \cite[Eq. (17)]{HsiehDevetakWinter:08p},
$%
    (U\otimes\text{id})\ket{\Phi_{AB}} = (\text{id}\otimes U^{T})\ket{\Phi_{AB}} %
$, %
we can then reflect the unitary operation to the entangled resource at the receiver along with the environment:
\begin{align}
    \ket{\chi^{\gamma,x^{n}}_{A^{n}G_2^{n}}} 
    &=(\identity\otimes U^{T}(\gamma))\ket{\psi^{x^{n}}_{A^{n}G_2^n}} \,.
    \label{apply_ricochet}
\end{align}

Thus, we can write the output state as follows:
\begin{align}
    \rho^{\gamma,x^{n}}_{B^{n}E^{n}G_2^{n}} &= (\mathcal{N}_{A\to BE}^{\otimes n}\otimes\mathrm{id})(\chi^{\gamma,x^{n}}_{A^{n}G_2^{n}})
    \nonumber\\
    &=
    (\mathcal{N}_{A\to BE}^{\otimes n}\otimes\mathrm{id})((\identity\otimes U^{T}(\gamma))\psi^{x^{n}}_{A^{n}G_2^n}(\identity\otimes U^{*}(\gamma)))
    \nonumber\\
    &=(\identity\otimes \identity\otimes U^{T}(\gamma))\left[(\mathcal{N}_{A\to BE}^{\otimes n}\otimes\mathrm{id})(\psi^{x^{n}}_{A^{n}G_2^n})\right](\identity\otimes \identity\otimes U^{*}(\gamma))
    \nonumber\\
    &= (\identity\otimes \identity\otimes U^{T}(\gamma))\omega^{x^{n}}_{B^{n}E^{n}G_2^n}(\identity\otimes \identity\otimes U^{*}(\gamma)) 
    \label{channel_output}
\end{align}
where $\omega^{x}_{BEG_2}$ is as  in \eqref{define_omega}.

Next, we analyze the error probability in each of our scenarios.

\vspace{0.1cm}
\subsubsection{Eve has stolen the resource}
We begin with the pessimistic case, where Bob does not have the  entangled resource $G_2^n$, as it was stolen by Eve. Bob's reduced state is given by
\begin{align}
    \rho_{B^n}^{\gamma, x^{n}} &= \trace_{E^{n}G_2^{n}}(\rho^{\gamma,x^{n}}_{B^{n}E^{n}G_2^{n}})
    \nonumber\\
    &= \trace_{E^{n}G_2^{n}}((\identity\otimes \identity\otimes U^{T}(\gamma))\omega^{x^{n}}_{B^{n}E^{n}G_2^n}(\identity\otimes \identity\otimes U^{*}(\gamma))) 
    \nonumber\\
    &= \omega^{x^{n}}_{B^{n}}
\end{align}
The second equality follows from trace cyclicity,  as $U^{*}U^{T}=\identity$.
Observe that the  state does not depend on $\gamma$. That is,
the reduced output state is not affected by
the encoding of $m'$. Therefore, based on the %
HSW Theorem \cite{Holevo:98p,SchumacherWestmoreland:97p}, %
 there exists a decoding POVM $\mathcal{D}^*_{B^n}=\{\Lambda_{m,k}\}$ such that
\begin{align}
    \mathbb{E}%
    P^{*(n)}_{e}(\Psi, \mathcal{F}, \mathcal{D}^*) 
    &\leq
    2^{-n[I(X;B)_\omega-R-R_0 - \epsilon_1]}
    \label{error_expectation_case_A}
\end{align}
for arbitrarily small $\epsilon_1$ and sufficiently large $n$, where the expectation is over the class of all codebooks.
Hence, if
\begin{align}
    R+R_0 < I(X;B)_\omega - \epsilon_1
    \label{demand_R_R_0}
\end{align}
then the expected error probability 
satisfies
\begin{align}
    \mathbb{E}%
    P^{*(n)}_{e}(\Psi, \mathcal{F}, \mathcal{D}^*)  \leq \alpha .
    \label{first_alpha_bound}
\end{align}

\vspace{0.1cm}
\subsubsection{Bob has entanglement assistance} 
We move to the optimistic case, where Eve has failed to intercept $G_2^n$, hence Bob holds  the entangled resource.
Based on the analysis above, Bob's first measurement recovers the correct guaranteed message $m$, with a high probability.
In general, upon performing a measurement,  it may lead to a state collapse. Denote the post-measurement state, after the first measurement, by $\tilde{\rho}_{B^nG_2^{n}}^{\gamma, x^{n}}$. According to the gentle measuring lemma \cite{Pereg:21p, Winter:99p, OgawaNagaoka:07p}, this state is close in trace distance to the original state, before the measurement took place, as 
\begin{align}
    \frac{1}{2}\norm{\tilde{\rho}_{B^nE^n G_2^{n}}^{\gamma, x^{n}}-\rho_{B^nE^n G_2^{n}}^{\gamma, x^{n}}} \leq 2^{-n\frac{1}{2}(I(X;B)_\omega -R -R_0- \epsilon_3)} \,,
\end{align}
which tends to zero by
 \eqref{demand_R_R_0}. Hence, we may focus  our error analysis on the original state, %
  before the measurement: %
 \begin{align}
    \rho_{B^nG_2^{n}}^{\gamma, x^{n}} &= \trace_{E^{n}}(\rho^{\gamma,x^{n}}_{B^{n}E^{n}G_2^{n}})
    \nonumber\\
    &= \trace_{E^{n}}((\identity\otimes \identity\otimes U^{T}(\gamma))\omega^{x^{n}}_{B^{n}E^{n}G_2^n}(\identity\otimes \identity\otimes U^{*}(\gamma))) 
    \nonumber\\
    &= (\identity\otimes U^{T}(\gamma)) \omega^{x^{n}}_{B^{n}G_2^{n}} (\identity\otimes U^{*}(\gamma))
    \label{Bob_state_with_entanglement}
\end{align}
where $\gamma\equiv \gamma (m',k'|x^{n}(m,k))$, and $    \omega^{x^{n}}_{B^{n}G_2^{n}} = \trace_{E^n}(\omega^{x^{n}}_{B^{n}E^{n}G_2^n})$.
Based on the same arguments as without secrecy \cite{pereg2023communication},
there exists a POVM $\{\Upsilon_{m',k'|x^n}\}$ such that the expected error probability is bounded by
\begin{align}
     \mathbb{E}%
     P^{n}_{e}(\Psi, \mathcal{F}, \mathcal{D})%
     \leq \alpha %
     \label{second_alpha_bound}
\end{align}
for sufficiently large $n$, provided that
\begin{align}
    R'+R'_0 < I(G_{2};B|X)_\omega -\epsilon_2 \,.
    \label{demand_R'_R'_0}
\end{align}

\subsection{Secrecy Analysis}
We note that secrecy is required whether Eve has intercepted Bob's entanglement resource $G_2^n$ or not.
Then, consider the extended state, %
\begin{align}
    \rho_{MM'E^{n}G_2^n}&= 
\frac{1}{2^{n(R+R')}}\sum_{m=1}^{2^{nR}}\sum_{m'=1}^{2^{nR'}}
\ketbra{m}_{M}\otimes \ketbra{m'}_{M'} \otimes \frac{1}{2^{n(R_0+R_0')}}\sum_{k=1}^{2^{nR_0}}\sum_{k'=1}^{2^{nR_0'}} \rho^{\gamma(m',k'|x^n(m,k)),x^n(m,k)}_{E^{n}G_2^{n}}
\end{align}
where $M$ and $M'$ are classical registers that store the guaranteed and excess messages, respectively, and
$\rho_{E^nG_2^{n}}^{\gamma, x^{n}} = \trace_{B^{n}}(\rho^{\gamma,x^{n}}_{B^{n}E^{n}G_2^{n}})$. %

In order to bound %
 the leakage as defined in \eqref{leakage_definiton} by $\delta$, 
 it suffices to prove the following bounds:
\begin{align}
&    I(M;E^{n}G_{2}^{n})_\rho \leq \frac{\delta}{2} \,,
    \label{first_secrecy_req}
\\%
&    I(M';E^{n}G_{2}^{n}|M)_\rho \leq \frac{\delta}{2} \,.
    \label{second_secrecy_req}
\end{align}
Consider Eve's joint state, of her output and the entanglement resource, which could be in her possession.
Similarly, as before, %
we  express Eve's joint state %
as
\begin{align}
    \rho_{E^nG_2^{n}}^{\gamma, x^{n}} &= \trace_{B^{n}}(\rho^{\gamma,x^{n}}_{B^{n}E^{n}G_2^{n}})
    \nonumber\\
    &= \trace_{B^{n}}((I\otimes I\otimes U^{T}(\gamma))\omega^{x^{n}}_{B^{n}E^{n}G_2^n}(I\otimes I\otimes U^{*}(\gamma))) 
    \nonumber\\
    &= (I\otimes U^{T}(\gamma)) \omega^{x^{n}}_{E^{n}G_2^{n}} (I\otimes U^{*}(\gamma))
    \label{rho_omega_eve}
\end{align}
where $  \omega^{x^{n}}_{E^{n}G_2^{n}} = \trace_{B^n}(\omega^{x^{n}}_{B^{n}E^{n}G_2^n})$ (see \eqref{define_omega}).

Next, we analyze the secrecy for each of Alice's messages.

\vspace{0.1cm}
\subsubsection{Guaranteed information leakage} %
Observe that if we were to remove the second encoding operation for the excess message, then Eve's output would have been
\begin{align}
    \widetilde{\omega}_{MM'E^{n}G_2^n}&= 
\frac{1}{2^{n(R+R')}}\sum_{m=1}^{2^{nR}}\sum_{m'=1}^{2^{nR'}}
\ketbra{m}_{M}\otimes \ketbra{m'}_{M'} \otimes \frac{1}{2^{n(R_0+R_0')}}\sum_{k=1}^{2^{nR_0}}\sum_{k'=1}^{2^{nR_0'}} \omega^{x^n(m,k)}_{E^{n}G_2^{n}} 
\label{define_omega_tilde}
\end{align}
(cf. \eqref{rho_omega_eve} and \eqref{define_omega_tilde}).
Therefore, by the data processing inequality,
it suffices to bound the leakage with respect to $\widetilde{\omega}_{ME^n G_2^n}$,
because
\begin{align}
I(M;E^n G_2^n)_\rho\leq I(M;E^n G_2^n)_{\widetilde{\omega}}
\label{Equation:Leakage_Guaranteed_Omega_tilde}
\end{align}
(see \cite[Theorem 11.9.3]{Wilde:17b}).
We apply the quantum covering lemma \cite{ahlswede2002strong},
Lemma~\ref{quantum_covering_lemma}, with the ensemble below,
\begin{align}
    \{p_{X^n}(x^n), \omega^{x^n}_{E^{n}G_{2}^{n}}\}_{x^n\in\mathcal{X}^n} \,,
\end{align}
and the following typical projectors,
\begin{align}
    &\Pi = \Pi_{\delta}^{(n)}(\omega_{E^{n}G_{2}^{n}}) \,,
\\
    &\Pi_{x^n} = \Pi_{\delta}^{(n)}(\omega_{E^{n}G_{2}^{n}}|x^n) \,,
\end{align}
(see Appendix~\ref{Appendix:Quantum_Method_of_Types} for the definition of the $\delta$-typical projectors). 
Based on standard typical projectors properties that are listed in Appendix~\ref{Appendix:Quantum_Method_of_Types}, %
there exists $\lambda>0$ such that 
\begin{align}
    \trace(\Pi\omega_{EG_{2}}^{\otimes n})&\geq 1-e^{-
    \lambda n} \,,
\\
    \trace(\Pi_{x^n}\omega_{EG_{2}}^{\otimes n})&\geq 1-
    e^{-\lambda n} \,,
\\
    \trace(\Pi)&\leq 2^{n(H(EG_{2})_\omega+\epsilon_{4})} \,,
\\
    \Pi_{x^n} \omega^{x^n}_{E^{n}G_{2}^{n}} \Pi_{x^n} &\leq 2^{-n(H(EG_{2}|X)_\omega-\epsilon_{4})} \Pi_{x^n} \,.
    \label{condition_4_pacing_lemma}
\end{align}
Thus, by Lemma~\ref{quantum_covering_lemma},
for every $ m\in [1:2^{nR}]$ and sufficiently large $n$,
 \begin{align}
     \Pr\left(\norm{\frac{1}{2^{nR_0}}\sum_{k=1}^{2^{nR_0}} \omega^{X^{n}(m,k)}_{E^{n}G_{2}^{n}}-\omega_{E G_{2}}^{\otimes n}}_1 >e^{-\frac{\lambda }{2}n} \right)
     \leq %
     \exp{-%
     2^{n(R_0-I(X;EG_2)_\omega-\epsilon_5)}} \,. %
     \label{prob_sec_1}
 \end{align}
The last bound tends to zero in a double exponential rate, provided that
 \begin{align}
     R_0 > I(X;EG_{2})_\omega +\epsilon_{5} \,.
     \label{condition_R_0}
 \end{align}

Observe that since the bound in \eqref{prob_sec_1} is double exponential, it also follows that the probability that 
\begin{align}
\norm{\frac{1}{2^{nR_0}}\sum_{k=1}^{2^{nR_0}} \omega^{x^{n}(m,k)}_{E^{n}G_{2}^{n}}-\omega_{E G_{2}}^{\otimes n}}_1 >e^{-\frac{\lambda }{2}n} \,,\; \text{for some $m\in [1:2^{nR}]$}
\end{align}
also tends to zero, by the union bound.
Equivalently, the probability that 
\begin{align}
\norm{\frac{1}{2^{nR_0}}\sum_{k=1}^{2^{nR_0}} \omega^{x^{n}(m,k)}_{E^{n}G_{2}^{n}}-\omega_{E G_{2}}^{\otimes n}}_1 \leq e^{-\frac{\lambda }{2}n} \,,\; \text{for all $m\in [1:2^{nR}]$}
\end{align}
 tends to $1$.
The event above implies that
\begin{align}
\Delta_n&\equiv \norm{\widetilde{\omega}_{ME^{n}G_2^n}-\frac{\identity_M}{2^{nR}}\otimes \omega_{E G_{2}}^{\otimes n}}_1
\nonumber\\
&\leq \frac{1}{2^{nR}} \sum_{m=1}^{2^{nR}} \norm{\frac{1}{2^{nR_0}}\sum_{k=1}^{2^{nR_0}} \omega^{x^{n}(m,k)}_{E^{n}G_{2}^{n}}-\omega_{E G_{2}}^{\otimes n}}_1
\nonumber\\
&\leq e^{-\frac{\lambda }{2}n} \,,
\end{align}
where the first inequality follows from the triangle inequality.
 By the Alicki-Fanne-Winter (AFW) inequality \cite{AlickiFannes:04p, Winter:16p},
this, in turn, implies
\begin{align}
I(M;E^n G_2^n)_{\widetilde{\omega}}
&=
H(M)-H(M|E^n G_2^n)_{\widetilde{\omega}}
\nonumber\\
&\leq 
nR\Delta_n
+(1+\Delta_n)H_2\left( \frac{\Delta_n}{1+\Delta_n} \right)
\nonumber\\
&\leq 
nRe^{-\frac{\lambda }{2}n}
+(1+e^{-\frac{\lambda }{2}n})H_2\left( \frac{e^{-\frac{\lambda }{2}n}}{1+e^{-\frac{\lambda }{2}n}} \right)
\nonumber\\
&\leq 
\frac{\delta}{2}
\end{align}
for sufficiently large $n$, where $H_2(x)\equiv -x\log(x)-(1-x)\log(1-x)$ is the binary entropy function.
Therefore, by \eqref{prob_sec_1},
\begin{align}
\Pr(I(M;E^n G_2^n)_{\widetilde{\omega}} > \frac{\delta}{2})
\leq \exp{-%
     2^{n(R_0-I(X;EG_2)_\omega-\epsilon_5)}} \,.
     \label{secrecy_criterion_1}
\end{align}

\vspace{0.1cm}
\subsubsection{Excess information leakage} 
\label{Subsubsection:Excess_Leakage}%
Observe that
\begin{align}
 I(M';E^{n}G_{2}^{n}|MK)_\rho
&=H(M'|MK)-H(M'|M K E^{n}G_{2}^{n})_\rho
\nonumber\\
&\stackrel{(a)}{=} H(M')-H(M'|M K E^{n}G_{2}^{n})_\rho
\nonumber\\
&\stackrel{(b)}{\geq} H(M')-H(M'|M  E^{n}G_{2}^{n})_\rho
\nonumber\\
&\stackrel{(a)}{=} H(M'|M)-H(M'|M K E^{n}G_{2}^{n})_\rho
\\
&=I(M';E^{n}G_{2}^{n}|M)_\rho
\end{align}
where $K$ is a classical register that stores the uniform key $k\in [1:2^{nR_0}]$. The equalities labeled with $(a)$ hold since $M'$ is statistically independent of 
$(M,K)$, and $(b)$ holds since conditioning cannot increase entropy.
Thus, it suffices to bound the conditional leakage 
$I(M';E^{n}G_{2}^{n}|M=m,K=k)_\rho$, given $m$ and $k$.

Then, let $x^n\equiv x^n(m,k)$ be fixed. 
Consider the uniform ensemble,
\begin{align}
    \left\{p(\gamma|x^n)=\frac{1}{\abs{\Gamma_{x^n}}}, \rho^{\gamma, x^n}_{E^{n}G_{2}^{n}}\right\}_{\gamma\in \Gamma_{x^n}} \,.
    \\
    \intertext{Using the quantum covering lemma, we will show that Alice's encoding simulates the average state,}
    \zeta^{x^n}_{E^{n}G_{2}^{n}} = \frac{1}{\abs{\Gamma_{x^n}}} \sum_{\gamma\in\Gamma_{x^n}}\rho^{\gamma, x^n}_{E^{n}G_{2}^{n}} \,.
\end{align}

Define the code projectors in terms of the $\delta$-typical projectors and the encoding unitary:
\begin{align}
    \Pi = \Pi_{\delta}^{(n)}(\omega_{E^n}|x^n) \otimes \Pi_{\delta}^{(n)}(\omega_{G_{2}^n}|x^n) \,,
    \label{covering_operator_3}
\end{align}
\begin{align}
    \Pi_{\gamma} =(I\otimes U^{T}(\gamma)) \Pi_{\delta}^{(n)}(\omega_{E^nG_2^{n}}|x^n)(I\otimes U^{*}(\gamma)) \,,
    \label{covering_operator_4}
\end{align}
where
$%
    \omega_{E^n}
    $ and $ 
    \omega_{G_2^n}
$ are the reduced states of $\omega_{E^{n}G_{2}^{n}}$ (see Appendix~\ref{Appendix:Quantum_Method_of_Types} for the definition of the $\delta$-typical projectors). %
In order to apply the quantum covering lemma, we need 
to show that there exists $\mu >0$ such that
the following properties hold:
\begin{align}
    \trace(\Pi\rho^{\gamma, x^n}_{E^{n}G_{2}^{n}})&\geq 1-e^{-\mu n}
    \label{covering_property_1}
    \\
    \trace(\Pi_{\gamma}\rho^{\gamma, x^n}_{E^{n}G_{2}^{n}})&\geq 1-e^{-\mu n}
    \label{covering_property_2}
    \\
    \trace(\Pi)&\leq 2^{n(H(E|X)_\omega+H(G_2|X)_\omega+\epsilon_{6})}
    \label{covering_property_3}
    \\
    \Pi_{\gamma} \rho^{\gamma, x^n}_{E^{n}G_{2}^{n}} \Pi_{\gamma} &\leq 2^{-n(H(EG_{2}|X)_\omega-\epsilon_{6})} \Pi_{\gamma}
    \label{covering_property_4}
\end{align}
The proof for \eqref{covering_property_1}-\eqref{covering_property_3}
is given in Appendix~\ref{proof_conditions_covering}, as their derivation follows from the $\delta$-typical projector properties that are listed in Appendix~\ref{Appendix:Quantum_Method_of_Types}, along with  standard arguments.
Now, we show the last bound in \eqref{covering_property_4}.

Observe that
\begin{align}
&\Pi_{\gamma} \rho^{\gamma, x^n}_{E^{n}G_{2}^{n}} \Pi_{\gamma}
\nonumber\\
&=
(\identity\otimes U^T(\gamma))\Pi_{\delta}^{(n)}(\omega_{E^nG_2^{n}}|x^n)(\identity\otimes U^*(\gamma))(\identity\otimes U^T(\gamma))\omega^{x^n}_{E^nG_2^{n}}(\identity\otimes U^*(\gamma))(\identity\otimes U^T(\gamma))\Pi_{\delta}^{(n)}(\omega_{E^nG_2^{n}}|x^n)(\identity\otimes U^*(\gamma))
\nonumber\\
&=
(\identity\otimes U^T(\gamma))\cdot \Pi_{\delta}^{(n)}(\omega_{E^nG_2^{n}}|x^n)\omega^{x^n}_{E^nG_2^{n}}\Pi_{\delta}^{(n)}(\omega_{E^nG_2^{n}}|x^n)\cdot(\identity\otimes U^*(\gamma))
\end{align}
where the last equality holds since $U^T(\gamma)=(U^*(\gamma))^{-1}$.
Based on the  properties of conditional typical projectors,
\begin{align}
    \Pi_{\delta}^{(n)}(\omega_{E^nG_2^{n}}|x^n)\omega^{x^n}_{E^nG_2^{n}}\Pi_{\delta}^{(n)}(\omega_{E^nG_2^{n}}|x^n)\ 
    &\leq 2^{-n(H(EG_2|X)_\omega -\epsilon_6)}\Pi_{\delta}^{(n)}(\omega_{E^nG_2^{n}}|x^n)
\end{align}
(see  \eqref{projector_property_6} in Appendix~\ref{Appendix:Quantum_Method_of_Types}).
Thus,
\begin{align}
\Pi_{\gamma} \rho^{\gamma, x^n}_{E^{n}G_{2}^{n}} \Pi_{\gamma}
&\leq 2^{-n(H(EG_2|X)_\omega -\epsilon_6)}
(\identity\otimes U^T(\gamma)) \Pi_{\delta}^{(n)}(\omega_{E^nG_2^{n}}|x^n)(\identity\otimes U^*(\gamma))
\nonumber\\
&\equiv  2^{-n(H(EG_2|X)_\omega -\epsilon_6)}\Pi_\gamma
\end{align}
where the last equality follows the definition of $\Pi_\gamma$ in  \eqref{covering_operator_4}, thus proving
\eqref{covering_property_4}.

By Lemma~\ref{quantum_covering_lemma},
 for every $m'\in [1:2^{nR'}]$ and sufficiently large $n$, %
 \begin{align}
     &\Pr\left(\norm{\frac{1}{2^{nR'_0}}\sum_{k'=1}^{2^{nR_0'}} \rho^{\gamma(m',k'|x^n), x^{n}}_{E^{n}G_{2}^{n}}-\zeta^{x^n}_{E^{n}G_{2}^{n}}}_1 > e^{-\frac{\mu }{2} n} \right)
     \leq 
     \exp{-%
     2^{n(R'_0-I(G_2;E|X)_\omega%
     -\epsilon_{7})}} 
     \label{prob_sec_3}
 \end{align}
which tends to zero in a double exponential rate, provided that
 \begin{align}
     R'_0 > I(E;G_{2}|X)_\omega +\epsilon_{7} \,.
     \label{condition_R'_0}
 \end{align}

Similarly, as for the guaranteed information analysis,
 since the bound in \eqref{prob_sec_3} is double exponential,  the probability that 
\begin{align}
\norm{\frac{1}{2^{nR'_0}}\sum_{k=1}^{2^{nR_0}} \rho^{\gamma(m',k'|x^n), x^{n}}_{E^{n}G_{2}^{n}}-\zeta^{x^n}_{E^{n}G_{2}^{n}}}_1 \leq e^{-\frac{\mu }{2} n} \,,\; \text{for all $(m,k,m')\in [1:2^{nR'}]\times [1:2^{nR_0}]\times [1:2^{nR'}]$}
\label{Equation:Indistinguishibility_all_m_prime}
\end{align}
 tends to $1$.
The event in \eqref{Equation:Indistinguishibility_all_m_prime}, along with
  the AFW inequality \cite{AlickiFannes:04p, Winter:16p},
 imply
\begin{align}
I(M';E^{n}G_{2}^{n}|MK)_\rho
&=
H(M')-H(M'|E^{n}G_{2}^{n}MK)_{\rho}
\nonumber\\
&\leq 
nR' e^{-\frac{\mu }{2}n}
+(1+e^{-\frac{\mu }{2}n})H_2\left( \frac{e^{-\frac{\mu }{2}n}}{1+e^{-\frac{\mu }{2}n}} \right)
\nonumber\\
&\leq 
\frac{\delta}{2}
\end{align}
for sufficiently large $n$.
Therefore, by \eqref{prob_sec_3},
\begin{align}
\Pr(I(M';E^{n}G_{2}^{n}|MK)_\rho > \frac{\delta}{2})
\leq \exp{-%
     2^{n(R'_0-I(G_2;E|X)_\omega%
     -\epsilon_{7})}} 
     \,.
     \label{secrecy_criterion_2}
\end{align}

\subsubsection{Deterministic codebook}
To complete the proof, we use the standard arguments in order to show that there exists a deterministic codebook 
$%
\big\{x^{n}(m,k), \gamma(m',k'|x^n(m,k)) \big\}
$, %
such that both  error probabilities and both leakages vanish.
Consider the following  error events,
\begin{align}
    \mathscr{A}_0 &= \{P^{*(n)}_{e}(\Psi, \mathcal{F}, \mathcal{D}^*)>\epsilon\} \,,
\\
    \mathscr{A}_ 1 &= \{P^{(n)}_{e}(\Psi, \mathcal{F}, \mathcal{D})>\epsilon\} \,,
\\
     \mathscr{B}_0 &=\{I(M;E^{n}G_{2}^{n})_\rho> \frac{\delta}{2} \} \,,
    \label{first_secrecy_demand}
\\
   \mathscr{B}_1 &=\{ I(M';E^{n}G_{2}^{n}|M)_\rho> \frac{\delta}{2} \} \,,
    \label{second_secrecy_demand}
\end{align}
where $0<\epsilon<1$.
By the union bound,
\begin{align}
    \Pr(\mathscr{A}_0\cup \mathscr{A}_1 \cup \mathscr{B}_0 \cup \mathscr{B}_1) \leq
    P(\mathscr{A}_0)+P(\mathscr{A}_1)+P(\mathscr{B}_0)+P(\mathscr{B}_1)
    \label{prob_union_bound}
\end{align}

Recall that by \eqref{first_alpha_bound} and \eqref{demand_R'_R'_0},  the expected value of both error probabilities is bounded by $\alpha$, where $\alpha$ is arbitrarily small. %
Choosing
$\alpha = \frac{1}{4}\epsilon ^2$ , we have by Markov's inequality,
\begin{align}
    \Pr(\mathscr{A}_0) = \Pr(P^{*(n)}_{e}(\Psi, \mathcal{F}, \mathcal{D}^*)>\epsilon)\leq 
    \frac{\mathbb{E}%
P^{*(n)}_{e}(\Psi, \mathcal{F}, \mathcal{D}^*)}{\epsilon} \leq \frac{1}{4}\epsilon \,.
\end{align}   
Similarly, the second term on the r.h.s. of \eqref{prob_union_bound} is bounded by $\frac{1}{4}\epsilon$ as well.
The third term tends to zero by 
\eqref{Equation:Leakage_Guaranteed_Omega_tilde}
and
\eqref{secrecy_criterion_1}, while the fourth tends to zero by 
\eqref{secrecy_criterion_2}.
We deduce that there exists
a $(2^{nR}, 2^{nR'}, n, \epsilon,\delta)$  secrecy code with unreliable entanglement assistance, if
\begin{align*}
    R+R_0 &< I(X;B)_\omega - \epsilon_1 \,,
    \\
    R'+R'_0 &< I(G_{2};B|X)_\omega -\epsilon_2 \,,
    \\
    R_0 &> I(EG_{2};X)_\omega +\epsilon_5 \,,
    \\
    R'_0 &> I(E;G_{2}|X)_\omega +\epsilon_7 \,,
\end{align*}
which reduces to
\begin{align*}
    R&< I(X;B)_\omega - I(X;EG_2)_\omega - \epsilon_1-\epsilon_5 \,,
    \\
    R'&< I(G_2;B|X)_\omega-I(G_2;E|X)_\omega -\epsilon_2-\epsilon_7 \,.
\end{align*}
This completes the achievability proof. 
\qed

\section{Proof of Theorem~\ref{Theorem:Degraded_Channels}}
\label{Proof_Converse}
Consider secure communication with unreliable entanglement assistance, and assume that the quantum wiretap channel $\mathcal{N}_{A\to B}$ is degraded. 
Suppose Alice and Bob would like to share the entangled resource $\Psi_{G^{n}_A G^{n}_B}$, yet Bob's share may be stolen by Eve. In our model, there are two scenarios. Namely, either
Bob holds the entanglement resource $G_B^n$, or Eve, depending on whether Eve has succeeded in her attempt to steal the resource. Alice first prepares a classical maximally correlated state,
\begin{align}
    \pi_{KMK'M'} &= 
    \left(
    \frac{1}{2^{nR}} \sum_{m=1}^{2^{nR}}\ketbra{m}_M\otimes\ketbra{m}_K
    \right) \otimes 
    \left(
    \frac{1}{2^{nR'}}\sum_{m'=1}^{2^{nR'}}\ketbra{m'}_{M'}\otimes\ketbra{m'}_{K'} 
    \right)
    \label{randomness_maximally_correlated_states}
\end{align}
where $M$, $K$, $M'$, and $K'$ are classical registers, such that $M$ and $K$ are in  perfect (classical) correlation, and so are $M'$ and $K'$. Bob needs to recover the value $m$ in both cases, whether he holds the resource or Eve. Whereas, Bob need only recover $m'$, if he holds the resource. Security requires that both $m$ and $m'$ are hidden from Eve, whether she intercepted $G_B^n$ or not.

Alice applies an encoding map $\mathcal{F}_{MM'G^{n}_A \to A^n}$ on $MM'$ and her share of entanglement, $G^{n}_A$. Hence, the input state is
\begin{align}
    \sigma_{KK' A^n G^{n}_B} &= (\identity\otimes \mathcal{F}_{MM'G^{n}_A \to A^n} \otimes \identity)(\pi_{KMK'M'} \otimes \Psi_{G^{n}_A G^{n}_B}) \,,
\end{align}
and transmits $A^n$ through $n$ channel uses of the quantum wiretap channel $\mathcal{N}_{A\to BE}$.
The output state  is thus,
\begin{align}
    \omega_{KK' B^n E^n G^{n}_B} &= (\mathrm{id}_{KK'} \otimes \mathcal{N}_{A\to BE} \otimes \mathrm{id}_{G_B^n})(\psi_{KK' A^n G^{n}_B}) \,.
\end{align}
If the entanglement resource is available to Bob, then he applies a decoding channel $\mathcal{D}_{B^n G_B^n\to \hat{M}\hat{M}'}$, creating
\begin{align}
\rho_{KK'\hat{M}\hat{M}' E^n}
 &= (\mathrm{id}_{KK'} \otimes \mathcal{D}_{B^nG^{n}_B \to \hat{M}\hat{M}'} \otimes \mathrm{id}_{E^n})(\omega_{KK' B^n G^{n}_B E^n}) \,.
\end{align}
If Eve has stolen the entanglement resource,  then Bob applies a decoding channel $\mathcal{D}^*_{B^n\to \tilde{M}}$, hence
\begin{align}
\rho^*_{KK'\tilde{M} E^n}
 &= (\mathrm{id}_{KK'} \otimes \mathcal{D}^*_{B^n \to \tilde{M}} \otimes \mathrm{id}_{G_B^n E^n})(\omega_{KK' B^n G^{n}_B E^n}) \,.
\label{Equation:Estimation_Pessimistic}
\end{align}

Consider a sequence of $(2^{nR},2^{nR'},n)$ codes, with vanishing errors and leakage, i.e.,
\begin{align}
\frac{1}{2}\norm{
\rho_{K\hat{M} K'\hat{M}'}
- \pi_{KMK'M'}
}_1 &\leq \alpha_n \,,
\label{Equation:Converse_Optimistic_Error}
\\
\frac{1}{2}\norm{
\rho^*_{K\tilde{M} }
- \pi_{KM}
}_1 &\leq \alpha_n^* \,,
\label{Equation:Converse_Pessimistic_Error}
\intertext{and}
I(KK';E^nG^{n}_B)_\omega &\leq \beta_n
\label{Equation:Converse_Leakage}
\end{align}
where $\alpha_n$, $\alpha_n^*$, and $\beta_n$ tend to zero as $n\to\infty$.
Based on entropy continuity, it also follows that %
\begin{align}
    \abs{I(K;M)_\pi-I(K;\hat{M})_{\rho^*}} &\leq n\varepsilon^*_n
    \label{demand_converse_1}
    \\
    \abs{I(K;M'|K)_\pi-I(K;\hat{M}'|K)_{\rho}} &\leq n\varepsilon_n
    \label{demand_converse_3}
\end{align}
where $\varepsilon^*_n,\varepsilon^*_n \to 0$ when $n \to \infty$ (see \cite[App.C, part B]{pereg2023communication})

The multi-letter bound on the guaranteed rate is based on standard arguments. Indeed, %
consider the scenario where Bob receives $B^n$ alone, while Eve gets both $E^n$ and $G^{n}_B$.
 Now,
\begin{align}
    nR &= I(K;M)_\pi 
    \nonumber\\
    &\leq I(K;\hat{M})_{\rho^*} + n\alpha^*_n 
    \nonumber\\
    &\leq I(K;B^n)_\omega+ n\alpha^*_n 
    \nonumber\\
    &\leq I(K;B^n)_\omega- I(K;E^nG^{n}_B)_\omega + n(\varepsilon_n^*+\beta_n)
    \label{converse_result_1}
\end{align}
where
the first inequality is due to \eqref{demand_converse_1}, the second inequality follows from the data processing inequality (see \eqref{Equation:Estimation_Pessimistic})
By \eqref{Equation:Converse_Leakage}, we have $I(K;E^n G_B^n)_\omega\leq \beta_n$. 
Note that the bound on $R$ does not require the degradedness property.

We move to the more challenging bound, on the excess rate. %
Here, we will use the degraded property of our quantum wiretap channel.
consider the scenario where Bob holds both $B^n$ and $G_B^n$.
As before, we use \eqref{Equation:Converse_Leakage}, \eqref{demand_converse_3}, and the data processing inequality to derive the following bound,
\begin{align}
    nR' &= I(K';M'|K)_\pi 
    \nonumber\\
    &\leq I(K';G^{n}_B B^n|K)_\omega - I(K';E^nG^{n}_B|K)_\omega +n(\varepsilon_n+\beta_n)
    \,.
\end{align}
By adding and subtracting the mutual information term $I(K';G_B^n|K)_\omega$, we can also write  this as
\begin{align}
    n(R'-\varepsilon_n-\beta_n) &\leq
    I(K';B^n|G^{n}_B K)_\omega - I(K';E^n|G^{n}_B K)_\omega 
    \nonumber\\
    &= I(K' G^{n}_B;B^n|K)_\omega - I(K' G^{n}_B;E^n|K)_\omega -\left[I(G^{n}_B;B^n|K)_\omega - I(G^{n}_B;E^n|K)_\omega\right]
    \label{converse_result_2}
\end{align}
by the chain rule for the quantum mutual information.
Due to our assumption that the quantum wiretap channel is degraded, the expression within the square brackets above is nonnegative. Thus, 
\begin{align}
n(R'-\varepsilon_n-\beta_n)
    &\leq I(K' G^{n}_B;B^n|K)_\omega - I(K' G^{n}_B;E^n|K)_\omega 
    \label{converse_result_3}
\end{align}
To complete the regularized converse proof, set $X = K$ and $G_2 =(K', G^{n}_B)$ in \eqref{converse_result_1} and \eqref{converse_result_3}, and take the limit of $n\to\infty$.
\qed

\section{Summary and Discussion}
\label{Section:Discussion}
We  study secure communication with unreliable entanglement assistance. Alice wishes to send a secret message to Bob, while exploiting pre-shared entanglement assistance. In our setting, the assistance is \emph{unreliable} as Eve may steal the entanglement resource  intended for Bob. %
Our present work continues the line of research that started with  \cite{pereg2023communication} and \cite{pereg2023communication-1b} on unreliable entanglement assistance. However, the previous work \cite{pereg2023communication,pereg2023communication-1b} did not include security concerns. 
A straightforward approach to leverage entanglement assistance is to generate a shared 
key, and then encode the information using the one-time pad protocol (see Remark~\ref{Remark:Shared_Key}). However, this strategy poses a security risk in our case. If Eve  intercepts the entanglement resource, then she will get a hold of Alice's  key, resulting in a failure and a breach of security. 

\subsection{Interception}
Our model involves possible interception of the entanglement resource.
Before communication begins, the legitimate parties try to generate entanglement assistance. To this end, Alice prepares an entangled pair of particles locally, and transmits one of them. While the particle travels from the transmitter, Eve tries to steal it. 
In the optimistic case, Eve fails. Hence, Alice and Bob have pre-shared entanglement resources, $G_A^{n}$ and $G_B^{n}$, respectively.
That is,
entanglement is successfully generated prior to the transmission of information, and Bob can decode while using his access to $G_B^{n}$  (see Figure~\ref{figure_switch}(a)).
Otherwise, in the pessimistic case, Eve intercepts the entanglement resource $G_B^{n}$, in which case, Bob must decode without it.  In other words, the entanglement is shared between Alice and Eve, instead of Bob (see Figure~\ref{figure_switch}(b)).
Nonetheless, despite the entangled resource being stolen by Eve, secrecy needs to be maintained.

In the communication phase, Alice remains unaware of whether Bob or Eve possesses the entanglement resource. 
However, based on the common use of heralded entanglement generation in practical systems \cite{barz2010heralded}, we posit the assumption that both Bob and Eve are aware of the recipient of $G_B^{n}$ (see also \cite[Remark 2]{pereg2023communication}).

 Eve's interception has severe consequences on entanglement-assisted communication \cite{QiSharmaWilde:18p} (see Remark~\ref{Remark:Superdense_Coding}). 
For example, 
suppose that Alice uses the super dense coding protocol to encode two classical bits, and then transmits her qubit  via a quantum erasure channel.
Consider the event that Bob receives an erasure, hence Eve receives the transmitted qubit.
Nevertheless, without the entanglement resource, there is no leakage, because each qubit by itself has no correlation with Alice's messages. On the other hand, if Eve has both qubits, then she can use the super dense decoder in order to recover Alice's bits.

\subsection{Coding With Unreliable Assistance}
In our
coding scheme, %
Alice encodes two messages, a guaranteed message  and an excess message, at  rates $R$ and $R'$, respectively, as she is uncertain whether
 it is Bob or Eve that holds the entanglement resource.
Bob is required to decode the guaranteed message, regardless of the location of the entanglement resource. If Bob has access to the entanglement resource, as intended by the legitimate parties, then he also decodes the excess message. 
In our  model, the communication is subject to 
 a strong secrecy constraint such that
 the messages are hidden from the eavesdropper, even if she intercepted the entanglement resource successfully, from her perspective. %

Our model considers two extreme scenarios, i.e.,
 the entanglement resources are either entirely available to Bob or not at all (see Remark~\ref{Remark:Hard_Decicion}). %
In digital communications, this strategy aligns with a \emph{hard decision} approach \cite{Proakis:01b}. Indeed, the decoder in our setting makes a hard decision on whether the  entanglement resources are viable. 
This approach fundamentally differs from noisy entanglement models that ensure reliability with respect to the average state.
 Qi et al. \cite{QiSharmaWilde:18p} consider entanglement-assisted communication via a quantum wiretap channel,  assuming that the eavesdropper has access to the legitimate receiver's entanglement resource. %
While their assumption that both Bob and Eve can measure the same system may seem to contradict the no-cloning theorem, the scenario that we have described above provides operational meaning to their assumption.

\subsection{Entanglement With Eve}
\label{Subsection:Discussion_Eve_Entanglement}
As opposed to traditional privacy models, in our model Eve is not completely passive, as she may capture the entanglement resource. This fundamental difference adds another layer of depth to our model and its characterization. %
As observed in Remark~\ref{Remark:Guaranteed_Message_Correlation}, the guaranteed information could have correlation with the receiver's entanglement resource.
Indeed, the guaranteed information $m$ needs to be encoded in such a manner that Bob could recover it even in the absence of the entanglement resource, see Figure~\ref{figure_switch}(b). 
Nevertheless,   Alice encodes \emph{her} resource $G_{A}^{n}$ using an encoding map that depends on both $m$ and $m'$ (see \eqref{encoding_messages}).
 As a result, the encoding operation may induce correlation between the guaranteed information $m$ and the entangled resource $G_{B}^{n}$.
 We have also seen the consequences of this observation on the rate region formula in Section~\ref{Section:Results}. %
 
While the setting resembles layered secrecy broadcast models \cite{zou2015broadcast, pereg2021key}, the analysis is much more involved, and the formulas have a different form (see Remark~\ref{Remark:Strange_Form}).
Specifically, instead of the mutual information term 
$I(X;E)_\omega$ in the private information formula, %
we now have $I(X;EG_2)_\omega$ that includes the receiver's entanglement resource, cf.
  \eqref{definition_private_information} and \eqref{R_S-EA*}.
Similarly, based on the model description,
 it may seem at a first glance as if $X$ should not be correlated with $G_2$, since the guaranteed information needs to be recovered in the absence of the entanglement resource (see Remark~\ref{Remark:Formula_Correlation_XG2}).  
However, as
mentioned above, %
   Alice's encoding %
   may induce correlation between the guaranteed information and the receiver's resource. %
It can also be seen in the rate region formula, where the application of the encoding channel $\mathcal{F}_{G_1\to A}^{(x)}$ could create correlation between $X$ and $G_2$ (see \eqref{Equation:omegaXG2A}).

\subsection{Single Letterization}
Deriving single-letter capacity formulas in notoriously difficult in quantum Shannon theory
(see \cite[Sec. IV-B]{pereg2023communication-1b}).
Given reliable entanglement assistance, Qi et al. \cite{QiSharmaWilde:18p} established a single-letter formula for the special class of degraded wiretap channels. 
Unfortunately,
in our model of unreliable entanglement assistance, we only have a multi-letter formula for degraded channels.   The difficulty in obtaining a single-letter formula arises from the unusual form of our rate region, 
as explained in Subsection~\ref{Subsection:Discussion_Eve_Entanglement} above.
Indeed, it is not clear whether we can restrict the encoder such that the channel from $A$ to $(E,G_2)$ (the combined output of Eve and the entanglement resource) is degraded.  
Establishing a bound on the dimension of the reference system in the secrecy capacity formula is an open problem, even with reliable entanglement assistance (see also \cite[Remark. 5]{QiSharmaWilde:18p} and \cite{PeregDeppeBoche:21p}).
Since the optimization is not being constrained to a pure state, the conventional approach based on the Schmidt decomposition  is not applicable   (see \cite{YardHaydenDevetak:08p, pereg2023communication}).

\subsection{Quantum Superposition Coding}
The analysis modifies the quantum superposition coding (SPC) scheme from \cite{pereg2023communication}.
The quantum SPC scheme
was inspired by the classical SPC technique  \cite{Cover:72p, CoverThomas:06b}. 
Originally, the classical SPC scheme consists of a collection of sequences $u^n(m)$ and $v^n(m')$, where $m$ and $m'$ are messages that are associated with different users in a multi-user network. 
In this scheme, the sequences $u^n(m)$ are referred to as cloud centers, while $v^n(m')$ are displacement vectors. The resulting codewords, denoted as$c^n(m,m')=u^n(m)+v^n(m')$ are referred to as satellites.

\begin{figure}[tbp]
 \centering
        \includegraphics[height=5cm, trim={0 0 0 0}, clip]{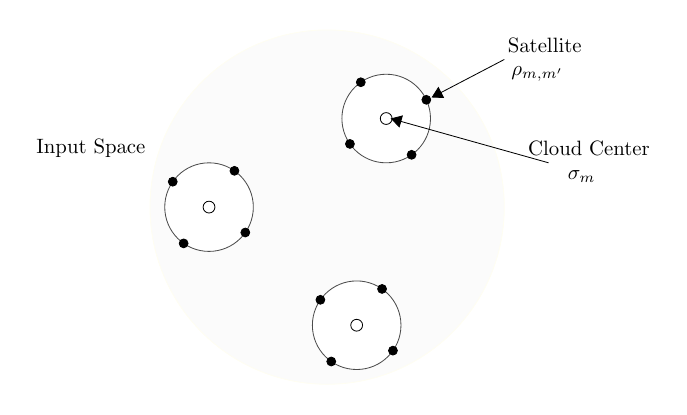} %
    \caption{Quantum superposition coding
    }
    \label{figure_SPC}
\end{figure}

In analogy, quantum SPC \cite{pereg2023communication} uses quantum operations that map quantum cloud centers to quantum satellite states. Suppose that Alice and Bob share an entangled state $\phi$ a priori. Each cloud center is associated with a classical sequence $x^n(m)$, and at the center of each cloud there is a state, $\sigma_m=\mathcal{F}^{(x^n(m))}(\phi)$, where $\mathcal{F}^{(x^n(m))}$ is a quantum encoding map that is conditioned on $x^n(m)$. Applying random Pauli operators that encode the message $m'$ takes us from the cloud center to a satellite $\rho_{m,m'}$ on the cloud that depends on both messages, $m$ and $m'$. The channel input is thus the satellite state. See Figure~\ref{figure_SPC}. Bob decodes in two steps. 
Initially, Bob aims to recover the cloud, i.e., he estimates the message $m$. If the entanglement resource is unavailable to Bob at this stage, the decoding process concludes after the first step. However, if Bob has access to entanglement assistance, he proceeds to the second step to decode the satellite associated with message $m'$. It was later shown that quantum SPC is optimal for entanglement-breaking quantum channels with unreliable entanglement assistance \cite{pereg2023communication-1b}. 

In our model, Alice inserts local randomness elements $k$ and $k'$ to confuse Eve.
Effectively, she
encodes the pair $(m,k)$ and $(m',k')$, instead of $m$ and   $m'$, respectively. Hence, the analysis in the secure setting is more involved compared to the basic quantum SPC from \cite{pereg2023communication}.

\subsection{Without Interception}
One may also consider an alternative model, where
Eve cannot intercept and steal the entanglement resource. In this case, if Alice and Bob generate entanglement between them successfully, then it can be used in order to generate a shared secret key. Therefore, the excess information $m'$ can be encoded using the one-time pad protocol.
This results in the following achievable region,
\begin{align}
\widetilde{\mathcal{R}}%
&\equiv
\bigcup_{ p_X, \varphi_{G_{1}G_{2}}, \mathcal{F}^{(x)} } 
\left\{ \begin{array}{rl}
  (R,R') \,:\;
    R   \leq    &    [I(X;B)_\omega - I(X;E)_\omega]_+  \\
    R'  \leq    &    I(G_{2};B|X)_\omega \\
	\end{array}
\right\} \ 
\label{R_S-EA*-No-Interception}
\end{align}
where the union is as in \eqref{R_S-EA*}.
The bounds on $R$ and $R'$ are both simpler (cf. \eqref{R_S-EA*} and \eqref{R_S-EA*-No-Interception}).
In the guaranteed rate bound, we now have the standard mutual information term
$I(X;E)_\omega$, instead of $I(X;EG_2)_\omega$, as in \eqref{R_S-EA*}.
Furthermore, the term $I(G_2;E|X)_\omega$ need not be substracted from the excess information rate $R'$, due to the availability of a secret key in the optimistic scenario. 

\section*{Acknowledgments}
U. Pereg and M. Lederman were supported by  Israel Science Foundation (ISF), Grants 939/23 and 2691/23, German-Israeli Project Cooperation (DIP), Grant
2032991, and  Nevet Program of the Helen Diller Quantum Center at the Technion.
U. Pereg was also supported by the Israel VATAT Junior Faculty Program for Quantum Science and Technology through Grant 86636903, and the Chaya Career Advancement Chair, Grant 8776026.

\renewcommand{\theequation}{\thesection.\arabic{equation}}
\begin{appendices} %
{

\section{Typical Projectors}
\label{Appendix:Quantum_Method_of_Types}

We begin with the definition of a classical  type. The type of a classical sequence $x^n\in \mathcal{X}^{n}$ is defined as the empricial distribution  $\hat{P}_{x^n}(a) = \frac{N(a|x^n)}{n}$ for $ a\in \mathcal{X}$, where $N(a|x^n)$ is the number of occurences of the letter $a$ in the sequence $x^n$. 
Let $\mathcal{T}(\hat{P})$ denote the subset of $\mathcal{X}^n$ that consists of all sequences of  type $\hat{P}$.  %
The subset $\mathcal{T}(\hat{P})$ is called the type class of $\hat{P}%
$. 
Let $\delta>0$. 
The $\delta$-typical set, $\mathcal{A}_{\delta}^{(n)}(p_{X})$, with respect to a probability distribution $p_X$, is defined as the set of all sequences
$x^n$ such that $\abs{\hat{P}_{x^n}(a)-P_{X}(a)} \leq \delta p_X(a)$, for all $a\in\mathcal{X}$.

Next, we move to the quantum method of types. Consider an ensemble $\{p_{X}(x), \ket{x}\}_{x\in\mathcal{X}}$, with an average state, $\rho=\sum_{x\in\mathcal{X}}p_{X}(x)\ketbra{x}$. 
The $\delta$-typical projector with respect to the ensemple above projects onto the subspace that is spanned by $\ket{x^n}$, where $x^n$ are classical $\delta$-typical sequences, i.e.,
\begin{align}
    \Pi^{(n)}_{\delta}(\rho) = \sum_{x^n \in \mathcal{A}_\delta^{(n)}(p_X)} \ketbra{x^n} \,.
\end{align}

The typical projector satisfies the following properties.  
There exists $a>0$ and $\epsilon>0$ such that %
\begin{align}
    & 1-2^{-an}\leq \trace\{\Pi^{(n)}_{\delta}(\rho)\rho^{\otimes n}\} \leq 1 \,,
    \label{projector_property_1}
    \\
    &\trace\{\Pi^{(n)}_{\delta}(\rho)\} \leq 2^{n(1+\delta)H(\rho)}
    \,, \label{projector_property_2}
    \\
    &(1-2^{-an})2^{-n(1+\delta)H(\rho)}\Pi^{(n)}_{\delta}(\rho) \leq \Pi^{(n)}_{\delta}(\rho)\rho^{\otimes n} \Pi^{(n)}_{\delta}(\rho) \leq 2^{-n(1-\delta)H(\rho)}\Pi^{(n)}_{\delta}(\rho)
    \,,\label{projector_property_3}
\end{align}
for sufficiently large $n$ (see \cite[Th. 1.1]{Kramer:08n} and \cite[Sec. 15.1.2]{Wilde:17b}).

Furthermore, we now define  the conditional $\delta$-typical subspace and projector. 
Consider an ensemble $\{p_X(x),\rho_B^x\}$, with an average state
 $\sigma_B = \sum_{x\in\mathcal{X}}p_{X}(x)\rho^x_B$, and a spectral decomposition $\rho^x_ B = \sum_{y\in\mathcal{Y}}p_{Y|X}(y|x)\ketbra{\psi^{x,y}}$, for $x\in\mathcal{X}$. Given a fixed sequence $x^n\in\mathcal{X}^n$, define for every $a\in\mathcal{X}$, let 
 $I_n(a)$ denote the set of indices $i\in [1:n]$ such that
 $ x_i=a$. Then, define the conditional $\delta$-typical subspace %
 as the span of all states $\ket{\psi^{x^n,y^n}} = \bigotimes_{i=1}^{n}\ket{\psi^{x_i,y_i}}$, such that $y^{I_n(a)} \in \mathcal{A}_\delta^{(\abs{I_n(a)})}(p_{Y|X}(\cdot|a))$, for $a\in\mathcal{X}$.
Then, the conditional $\delta$-typical projector $\Pi^{(n)}_{\delta}(\sigma_B|x^n)$ %
is defined as projecting onto the conditional $\delta$-typical subspace.

Similarly to before,  the conditional typical projector 
satisfies the properties below. There exist $a>0$ and $\epsilon_n(\delta)$ such that 
\begin{align}
     &1-2^{-an}\leq \trace\{\Pi^{(n)}_{\delta}(\sigma_B|x^n)\rho_{B^n}^{x^n}\} \leq 1
    \label{projector_property_4}
    \\
    & \trace\{\Pi^{(n)}_{\delta}(\sigma_B|x^n)\} \leq 2^{n(1+\epsilon_n(\delta))H(B|X')_\sigma}
    \label{projector_property_5}
    \\
    & 2^{-n(1+\epsilon_n(\delta))H(B|X')_\sigma}\Pi^{(n)}_{\delta}(\sigma_B|x^n) \leq \Pi^{(n)}_{\delta}(\sigma_B|x^n)\rho_{B^n}^{x^n} \Pi^{(n)}_{\delta}(\sigma_B|x^n) \leq 2^{-n(1-\epsilon_n(\delta))H(B|X')_\sigma}\Pi^{(n)}_{\delta}(\sigma_B|x^n)
    \label{projector_property_6}
\end{align}
for sufficiently large $n$, where $\epsilon_n(\delta)$ tends to zero as $n\to\infty$ and $\delta\to 0$
(see \cite[Th. 1.2]{Kramer:08n} and \cite[Sec. 15.2.4]{Wilde:17b}),
 $\rho_{B^n}^{x^n} = \bigotimes_{i=1}^{n}\rho_{B_i}^{x_i}$, and the classical random variable $X'$ is distributed according to the type of $x^n$.

\section{Covering Lemma Properties}
\label{proof_conditions_covering}
We prove the first three properties, \eqref{covering_property_1}-\eqref{covering_property_3}, which are required to apply the quantum covering lemma in our secrecy analysis (see part~\ref{Subsubsection:Excess_Leakage} in the achievability proof in Section~\ref{Proof_Achievability}).
There arguments are similar to those in \cite[Appendix II]{HsiehDevetakWinter:08p}. For completeness, we give the details below.

Property \eqref{covering_property_3}  immediately follows from the dimension bound in \eqref{projector_property_5}.
As for property \eqref{covering_property_2}, we have
\begin{align}
    \trace(\rho^{\gamma, x^n}_{E^{n}G_{2}^{n}}\Pi_{\gamma})
    &=\trace((\identity\otimes U^T(\gamma)) \omega^{x^n}_{E^{n}G_{2}^{n}} (\identity\otimes U^*(\gamma)) (\identity\otimes U^T(\gamma)) \Pi_{\delta}^{(n)}(\omega_{E^nG_2^{n}}|x^n) (\identity\otimes U^T(\gamma)))
    \nonumber\\
    &= \trace((\identity\otimes U^T(\gamma)) \omega^{x^n}_{E^{n}G_{2}^{n}} \Pi_{\delta}^{(n)}(\omega_{E^nG_2^{n}}|x^n) (\identity\otimes U^T(\gamma)))
    \nonumber\\
    &= \trace(\omega^{x^n}_{E^{n}G_{2}^{n}} \Pi_{\delta}^{(n)}(\omega_{E^nG_2^{n}}|x^n))
    \nonumber\\
    &\geq 1-\frac{1}{2}\epsilon_6
\label{Equation:Covering1_Proof}
\end{align}
for sufficiently large $n$.
The first equality follows from substituting the expressions for 
$\rho^{\gamma, x^n}_{E^{n}G_{2}^{n}}$ and $\Pi_{\gamma}$ from
\eqref{rho_omega_eve} and \eqref{covering_operator_4}, respectively, the second equality holds since $ U^*(\gamma)= ( U^T(\gamma))^{-1}$, the third holds due to the trace cyclic property, and the inequality is due to the conditional typical projector property in \eqref{projector_property_4}. 

It remains to prove property \eqref{covering_property_1}. To this end, we express the projector $\Pi$ in terms of the complementary projectors below,
\begin{align}
\hat{\Pi}_{\delta}^{(n)}(\omega_{E^n}|x^n))&\equiv \identity-{\Pi}_{\delta}^{(n)}(\omega_{E^n}|x^n)) \,,
\label{Equation:Pi_Complenetary_E}
\\
\hat{\Pi}_{\delta}^{(n)}(\omega_{G_2^n}|x^n))&\equiv \identity-{\Pi}_{\delta}^{(n)}(\omega_{G_2^n}|x^n)) \,.
\label{Equation:Pi_Complenetary_G2}
\end{align}
Then,
\begin{align}
    \Pi 
    &= \Pi_{\delta}^{(n)}(\omega_{E^n}|x^n) \otimes \Pi_{\delta}^{(n)}(\omega_{G_{2}^n}|x^n)
    \nonumber\\
    &= (\identity - \hat{\Pi}_{\delta}^{(n)}(\omega_{E^n}|x^n)) \otimes (\identity - \hat{\Pi}_{\delta}^{(n)}(\omega_{G_2^n}|x^n))
    \nonumber\\
    &=
    (\identity\otimes\identity) - (\hat{\Pi}_{\delta}^{(n)}(\omega_{E^n}|x^n)\otimes\identity) - (\identity\otimes \hat{\Pi}_{\delta}^{(n)}(\omega_{G_2^n}|x^n)) + (\hat{\Pi}_{\delta}^{(n)}(\omega_{E^n}|x^n) \otimes \hat{\Pi}_{\delta}^{(n)}(\omega_{G_2^n}|x^n))
    \nonumber\\
    &\geq (\identity\otimes\identity) - (\hat{\Pi}_{\delta}^{(n)}(\omega_{E^n}|x^n)\otimes\identity) - (\identity\otimes \hat{\Pi}_{\delta}^{(n)}(\omega_{G_2^n}|x^n)) 
\end{align}
where the first equality holds by the definition of $\Pi$ in \eqref{covering_operator_3}, and the second by \eqref{Equation:Pi_Complenetary_E}-\eqref{Equation:Pi_Complenetary_G2}.

Therefore,
\begin{align}
    \trace(\rho^{\gamma| x^n}_{E^{n}G_{2}^{n}}\Pi) 
    &= \trace((\identity\otimes U^T(\gamma)) \omega^{x^n}_{E^{n}G_{2}^{n}} (\identity\otimes U^*(\gamma))\Pi)
    \nonumber\\
    &\geq \trace((\identity\otimes U^T(\gamma)) \omega^{x^n}_{E^{n}G_{2}^{n}} (\identity\otimes U^*(\gamma))(\identity\otimes\identity)) -\trace((\identity\otimes U^T(\gamma)) \omega^{x^n}_{E^{n}G_{2}^{n}} (\identity\otimes U^*(\gamma))(\hat{\Pi}_{\delta}^{(n)}(\omega_{E^n}|x^n)\otimes\identity))
    \nonumber\\
    &-\trace((\identity\otimes U^T(\gamma)) \omega^{x^n}_{E^{n}G_{2}^{n}} (\identity\otimes U^*(\gamma))(\identity\otimes \hat{\Pi}_{\delta}^{(n)}(\omega_{G_2^n}|x^n)))
    \label{app_cov_prop_2_1} \,.
\end{align}
By the trace cyclic property, the first term equals
$%
    \trace(\omega^{x^n}_{E^{n}G_{2}^{n}}) = 1
$, %
and the second equals
\begin{align}
    \trace(\omega^{x^n}_{E^{n}G_{2}^{n}}\hat{\Pi}_{\delta}^{(n)}(\omega_{E^n}|x^n))
    &= 1- \trace(\omega^{x^n}_{E^{n}G_{2}^{n}}\Pi_{\delta}^{(n)}(\omega_{E^n}|x^n))
    \nonumber\\
    &\leq %
    \frac{1}{2}
    \epsilon_6
\end{align}
where the inequality holds by \eqref{Equation:Covering1_Proof}.
Similarly, the last term is bounded by $\frac{1}{2}\epsilon_6$.
Substituting those terms in \eqref{app_cov_prop_2_1}
yields the desired bound \eqref{covering_property_1}.
\qed

}
\end{appendices}

\ifdefined\bibstar\else\newcommand{\bibstar}[1]{}\fi

\end{document}